\documentclass[a4paper,oneside,english]{aa}
\usepackage[T1]{fontenc}
\usepackage[koi8-r]{inputenc}
\usepackage{amsmath}
\usepackage{graphicx}
\usepackage{amssymb}

\makeatletter

\providecommand{\tabularnewline}{\\}

\usepackage{graphicx}
\usepackage{xspace}

\newcommand{\ascc}   {\mbox{ASCC-2.5}\xspace}
\newcommand{\clucat} {\mbox{COCD}\xspace}

\usepackage{babel}
\makeatother

\begin{document}
\titlerunning{Radii and mass segregation in open clusters} \authorrunning{Schilbach et al.}

\title{Population analysis of open clusters: radii and mass segregation}

\author{E.~Schilbach \inst{1} \and
        N.V.~Kharchenko \inst{1,2,4}\and
        A.E.~Piskunov \inst{1,3,4}\and
        S.~R\"{o}ser \inst{1}\and
        R.-D.~Scholz \inst{4} }

\offprints{R.-D.~Scholz}

\institute{Astronomisches Rechen-Institut, M\"{o}nchhofstra\ss{}e 12-14, D--69120
Heidelberg, Germany\\
 email: elena@ari.uni-heidelberg.de, nkhar@ari.uni-heidelberg.de,
apiskunov@ari.uni-heidelberg.de, roeser@ari.uni-heidelberg.de \and
           Main Astronomical Observatory, 27 Academica Zabolotnogo Str., 03680
Kiev, Ukraine\\
 email: nkhar@mao.kiev.ua \and
           Institute of Astronomy of the Russian Acad. Sci., 48 Pyatnitskaya
Str., Moscow 109017, Russia\\
 email: piskunov@inasan.rssi.ru \and
           Astrophysikalisches Institut Potsdam, An der Sternwarte 16, D--14482
Potsdam, Germany\\
 email: apiskunov@aip.de, nkharchenko@aip.de, rdscholz@aip.de }

\date{Received 8 December 2005; accepted 23 May 2006}

\abstract
%...Context
{}
%...Aims  
{Based on our well-determined sample of open clusters in the all-sky catalogue
\ascc we derive new linear sizes of some 600 clusters, and 
investigate the effect of mass segregation of stars in open clusters.} 
%...Methods
{Using statistical methods, we study
the distribution of linear sizes as a function of spatial position and
cluster age.
We also examine statistically the distribution of stars of different masses
within clusters as a function of the cluster age.}
%...Results
{No significant dependence of the cluster size on location in the Galaxy is
detected for younger clusters ($<$ 200~Myr), whereas older clusters inside the
solar orbit turned out to be, on average, smaller than outside. Also, small old
clusters are preferentially found close to the Galactic plane, whereas larger
ones more frequently live farther away from the plane and at larger
Galactocentric distances. For clusters with $(V - M_V) < 10.5$, a clear
dependence of the apparent radius on age has been detected: the cluster
radii decrease by a factor of about 2 from an age of 10 Myr to an age of 1 Gyr.
A detailed analysis shows that this observed effect can be explained by
mass segregation and does not necessarily reflect a real decrease
of cluster radii. We found evidence for the latter for the majority of clusters
older than 30 Myr. Among the youngest clusters (between 5 and 30 Myr),
there are some clusters with a significant grade of mass segregation, whereas
some others show no segregation at all. At a cluster age between
50 and 100 Myrs, the distribution of stars of different masses 
becomes more regular over cluster area. In older clusters the evolution of 
the massive stars is the
most prominent effect we observe.}
%...Conclusions
{}

\maketitle

\section{Introduction}

Open star clusters are gravitationally bound systems of, typically,
several hundreds of stars formed together. Primordial conditions during
the cluster formation and the location of the parental molecular cloud
in the Galaxy play an important role in the fate of a cluster. On
the other hand, the stellar content of a cluster evolves with time,
and internal and external interactions affect the properties
of individual cluster members as well as of the whole cluster as a system.
Therefore, the spatial structure and mass distribution that
we observe today in a given cluster is the result of the original
brand marks and the ongoing evolution.

Numerical simulations of the dynamical evolution predict a mass segregation
in open clusters i.e., a different concentration of cluster members
with different masses with respect to the cluster centre. This process
occurs on approximately the relaxation time-scale and is independent of
most of the possible initial conditions (Bonnell \& Davies~\cite{boda}).
During the dynamical evolution of a cluster, more massive members
sink to the centre, whereas less massive stars tend to show a diffuse 
distribution (Portegies Zwart \& McMillan~\cite{pzm}).
A relaxed cluster can be thought of as a set of nested spherical clouds
of stars of different mass (e.g., see Adams et al.~\cite{adea01}
for illustration). With increasing mass of the stars, the radial density
profile becomes steeper and narrower. Due to tidal interactions with
the Galactic gravitational field, the cluster can lose stars once
they overflow its tidal radius. Due to mass segregation a cluster
loses preferentially low mass stars from the cluster corona, which
evaporate into the field, up to an entire dissolution of the cluster
in the Galaxy (Andersen \& Nordstr\"{o}m~\cite{andea00}). A sudden
mass loss or close passage of giant molecular clouds can considerably
disturb a relaxation process and reduce the life-time of open clusters
(Kroupa et al.~\cite{krea01},
Bergond et al.~\cite{bea01}).

Predicted by simulations (Spitzer \& Shull~\cite{spish}), mass segregation
was already found in many open clusters. The most reliable results
on mass segregation can be expected for nearby clusters like the Pleiades
and Praesepe, where cluster members can be observed over a wide range
of magnitudes and masses. Compared to the distribution of more massive
stars, indications for a flatter density profile of cluster members
with $m<1\, m_{\odot}$ were obtained in these clusters by Jones \&
Stauffer~(\cite{jost91}), and more recently by Adams et al.~(\cite{adea01},
\cite{adea02}) who used data from 2MASS and USNO-A. Raboud
\& Mermilliod~(\cite{ramer98a}, \cite{ramer98b}) found evidence
for a continuous flattening of density profiles with decreasing mass
of cluster members in the Pleiades and Praesepe, and in a much more
distant open cluster, NGC 6231, too. Similar effects were detected
by Sagar et al.~(\cite{sagar88}), who considered 11 distant clusters
in the Galactic disk and by de Grijs et al.~(\cite{deg02a}, \cite{deg02b},
\cite{deg02c}), who studied mass segregation in open clusters in
the Large Magellanic Cloud.

Additionally to the mass segregation in older clusters due to dynamical
evolution, a higher concentration of massive stars to the centre was
also found in some very young clusters (e.g., in the Orion Nebula
Cluster, Hillenbrand~\cite{hi97}, Hillenbrand \& Hartmann~\cite{hiha}).
Due to the youth of these clusters, the central location of massive
stars can not be explained by dynamical evolution only. Additional
arguments from star formation scenarios and early cluster dynamics
have to be considered (Bonnell \& Davies~\cite{boda}, Kroupa
et al.~\cite{krea01}).

Since mass segregation has a direct impact on the spatial structure
of clusters, the effect should be seen in the {\it apparent} cluster sizes.
In this context, the observed cluster size is an important parameter related
to the dynamical state both of the cluster and of the Galactic disk.
As open clusters are found over a broad span of ages, a study of global
trends including cluster size, if they exist, should be possible from
a representative sample of clusters with homogeneous data on the main
cluster parameters.

Practically all famous collections (Trumpler, Collinder etc.) of cluster
data include estimations of angular sizes of open clusters, but the
first systematic determination of apparent diameters was made by Lyng\aa{}~(\cite{lyn87})
for about 1000 open clusters from visual inspection of the POSS prints.
These estimates are included in the Lund catalogue, 5th edition, together
with about other 150 clusters, where an estimate by G. Lyng\aa{}~
himself was not available, hence taken from the references quoted
in the catalogue. For reasons of homogeneity only the diameter of
the cluster nucleus (core) was included in the catalogue, although
already then it was known that some clusters showed coronae.
These data were used by Lyng\aa{}~(\cite{lyn82}) and Janes et al.~(\cite{janea88})
for their studies of properties of the open cluster system. Since
that time ages and distances were available for a small fraction of
known clusters, their sample included about 400 clusters. No estimations
were made on how well this sample represents the Galactic cluster population.

Although structural parameters have been derived for many individual
clusters during the last decade, there are only a few studies dealing
with a systematic determination of cluster sizes based on objective
and uniform approaches for larger cluster samples. Danilov \& Seleznev~(\cite{danil})
derived structural parameters for 103 compact distant ($>$ 1~kpc)
clusters from star counts down to $B\approx16$ from homogeneous wide-field
observations with a 50-cm Schmidt camera of the Ural university.
Based on \emph{UBV}-CCD observations compiled from literature, Tadross
et al.~(\cite{tad02}) redetermined ages and distances for 160 open
clusters. The cluster sizes were estimated visually, from POSS prints,
and they are practically identical to the diameters estimated by Lyng\aa{}.
Kharchenko et al.~(\cite{khea03}) determined radii of about 400 clusters
from star counts in \ascc and USNO-A2.0 catalogues.
Nilakshi et al.~(\cite{nilak}) derived structural parameters of
38 open clusters selected from the Lyng\aa's~(\cite{lyn87}) catalogue
from star counts in the USNO-A2.0 catalogue. Recently, Bonatto \&
Bica~(\cite{bonb}) published structural and dynamical parameters
of 11 open clusters obtained from star counts and photometric membership
based on the 2MASS survey.

Correlations of cluster size with age and Galactic location were found
by some of the authors above, though the results are rather controversial
(see \S~\ref{sec:location} and \S~\ref{sec:age} for more details).
There are at least two major aspects which must be taken into account
in the interpretation of the results. At first, how well does a given
sample represent the local population of open clusters in the Galaxy,
or which biases can arrise from the incompleteness of the data and
influence the results. Second, how homogeneous are data on individual
clusters, on their size, age, distance, provided that they are based
on observations with different telescopes equipped with different detectors,
or if different methods were used for the determination of cluster
parameters. The answer is not trivial considering the large set of
data compiled from literature, especially.

Using the Catalogues of Open Cluster Data (COCD%
\footnote{\texttt{ftp://cdsarc.u-strasbg.fr/pub/cats/J/A+A/438/1163,
ftp://cdsarc.u-strasbg.fr/pub/cats/J/A+A/440,403}%
}; Kharchenko et al.~\cite{starcat},~\cite{clucat},~\cite{newclu},
Paper I, II, III, respectively), we are able to reduce those uncertainties
which are due to the inhomogeneity of the cluster parameters, and
we can better estimate biases due to an incompleteness of the cluster
sample. The \clucat is originated from the All-Sky Compiled Catalogue
of 2.5 million stars (\ascc$\negthickspace$%
\footnote{\texttt{ftp://cdsarc.u-strasbg.fr/pub/cats/I/280A}%
}; Kharchenko~\cite{kha01}) including absolute proper motions in
the Hipparcos system, $B$, $V$ magnitudes in the Johnson photometric
system, and supplemented with spectral types and radial velocities
if available. The \ascc is complete up to about $V=11.5$~mag. We
identified 520 of about 1700 known clusters (Paper~I) in the \ascc and
found 130 new open clusters (Paper~III). Therefore, the completeness
of the cluster sample is mainly defined by the limiting magnitude
of the \ascc. For each cluster, membership was determined by use
of spatial, kinematic, and photometric information (Paper~I), and
a homogeneous set of structural, kinematic and evolutionary parameters
was obtained by applying a uniform technique (Papers II and III).
The sample was used to study the population of open clusters in the
local Galactic disk by jointly analysing the spatial and kinematic
distributions of clusters (Piskunov et al.~\cite{clupop}, Paper
IV).

In this paper we use the homogeneous data on structural parameters
of open clusters from the COCD to study general correlations including
cluster sizes as well as to analyse the spatial distribution of cluster
members from the point of view of mass segregation. In Sec.~\ref{sec:data}
we briefly describe the data set and estimate the statistic properties
of the cluster sample. The relations between cluster radius and its
location in the Galaxy are discussed in Sec.~\ref{sec:location}.
The correlations of cluster size with age is considered in Sec.~\ref{sec:age}.
In Sec.~\ref{sec:masseg} we examine the effect of mass segregation
in open clusters. A summary is given in Sec.~\ref{sec:concl}.

\section{Data\label{sec:data}}

The present study is based on the Catalogue of Open Cluster Data (\clucat)
and its Extension~1 described in Papers II and III. The complete sample consists
of 650 objects of which 641 are open clusters and 9 are compact associations.
For each \clucat object, the catalogue contains celestial position,
distance, reddening, age, angular sizes (core and cluster radii),
and kinematic data (proper motions and, if available, radial velocity). The cluster
parameters were obtained from a uniform data set (the \ascc catalogue)
and by use of a uniform technique of membership and parameter determination
(Papers I and III).

\begin{figure}
%\selectlanguage{russian}
\includegraphics[%
  bb=332bp 47bp 560bp 411bp,
  clip,
  height=0.95\linewidth,
  keepaspectratio,
  angle=270]{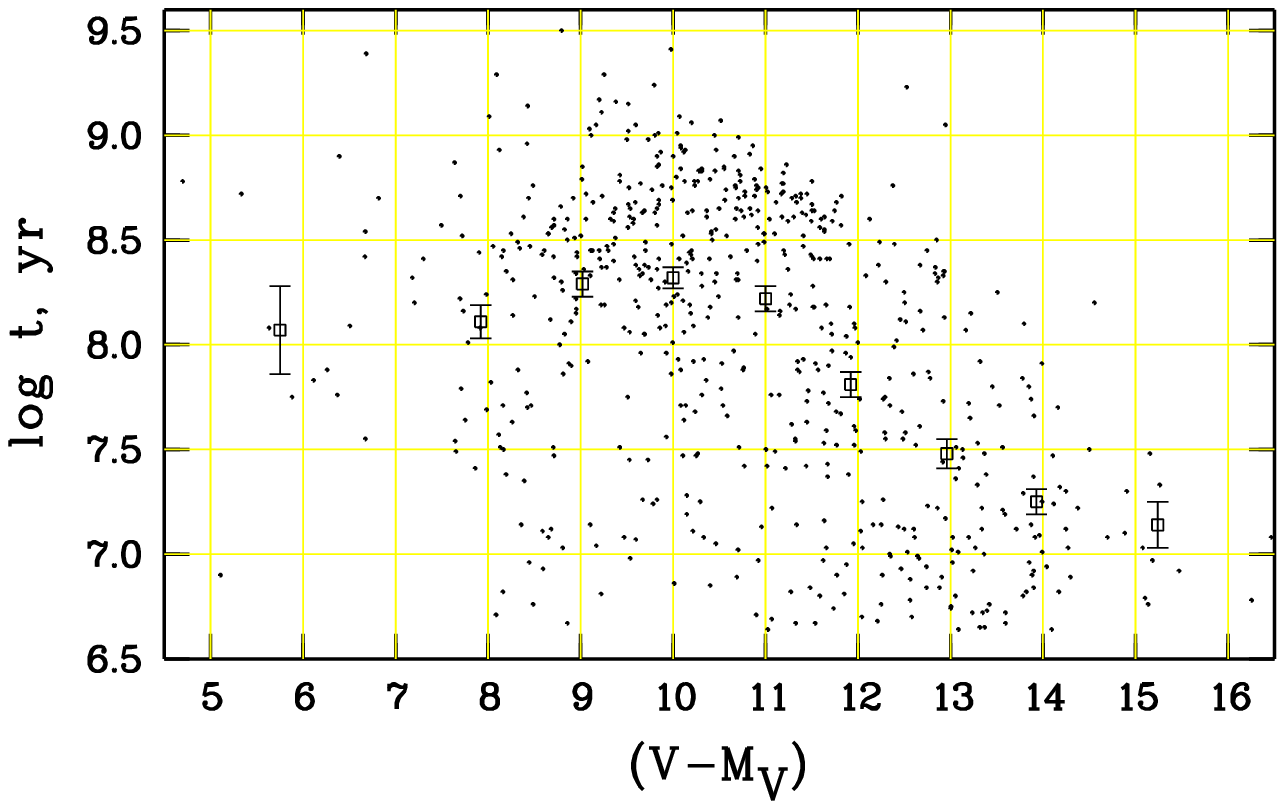}

\selectlanguage{english}
\caption{\label{fig:lgt-dmod}Cluster age versus distance modulus. Dots are
individual clusters. Their mean ages together with $r.m.s.$ errors
are shown as squares with bars.}
\end{figure}

\subsection{The completeness of the sample\label{sec:compl}}

In Paper IV we estimated the completeness limit of our cluster sample
to be 0.85~kpc. This result is based on the analysis of the surface
density of open clusters in the Galactic plane as a function of their
distance from the Sun. Nevertheless, for a statistical study of cluster
sizes the distance modulus $(V-M_{V})$ is, in many cases, a more
suitable parameter than a distance itself: due to the expected effect of 
mass segregation, more massive (i.e. more luminous) cluster members are 
concentrated to the cluster centre,  whereas fainter stars are located at the
cluster borders. Therefore, the apparent size
of a cluster depends on the brightness limit of the input catalogue, as well as
on the distance and extinction for a given cluster (i.e. on $(V-M_{V})$). 
Since reddening is known
for each cluster of our sample, we compute a typical distance
modulus $(V-M_{V})=5\log d-5+3.1\times E(B-V)$ for clusters at 0.85~kpc
from the Sun to be 10.0 - 10.5~mag. Taking into account
the completeness limit of the \ascc  at $V=11.5$, the corresponding
limit in absolute magnitudes of cluster members is about
1.5...1.0~mag. This absolute magnitude is still sufficiently faint
to observe MS stars in clusters younger than 1~Gyr, whereas older
clusters can be identified from their red giants. At distances larger
than the completeness limit we are steadily losing old clusters, and
our sample should get {}``younger'' on average. Indeed, the completeness
limit at $(V-M_{V})\approx10.0-10.5$ can be clearly concluded
from Fig.~\ref{fig:lgt-dmod} where we show the distribution of cluster
ages versus distance modulus.

\subsection{Linear sizes of open clusters\label{sec:size}}

For all clusters of our sample, we determined/redeter\-mined the
angular sizes by applying the same method (see Paper~II for details).
This approach is based on an iterative procedure which includes simultaneous
determination of membership and parameters for a cluster. The selection
of members takes into account photometric (location in the CMD) and
astrometric (proper motions and positions) criteria, and the standard
output parameters are the coordinates of the cluster centre, angular
size, mean proper motion, distance, extinction, and age.

Based on stellar counts, we considered two empirical 
structural components for each cluster, the core
and the corona (Paper~II). The distribution of $1\sigma$-members (i.e. stars
with the membership probability $P\geq61$\%) was the most important
factor for the determination of the cluster radius. The core radius
was defined as the distance from the cluster centre where the decrease
of stellar surface density gets flatter. The corona radius (or simply,
cluster radius) corresponds to a distance where the surface density
of stars becomes equal to the average density of the surrounding field.

The linear radii were computed from the individual distances and angular
sizes of the clusters. General properties of their distribution are
given in Table~\ref{tbl:rad}. For about 500 open clusters of our
sample, the sizes are also given in Lyng\aa{}~(\cite{lyn87}). As
expected (see also Paper~II), the cluster radii from Lyng\aa{}~(\cite{lyn87})
are in average lower by a factor of 2, and they fit rather the core
than the corona. Here we would like to stress the main advantage of
our approach with respect to other methods: our estimations of cluster
sizes are based on complete information on cluster membership which
includes the photometric as well as astrometric criteria. In order
to illustrate, as an example, the importance of reliable membership
data for a cluster parameter determination, we refer the reader to
the cluster Ruprecht 147 (cf. COCD, Atlas, page 460)
which is included in the catalogue of Lyng\aa{}~(\cite{lyn87}).
Neither distance nor age was known for this cluster, and a diameter
of 47~arcmin was estimated. Applying our procedure of membership
and parameter determination, we found that the cluster has a proper
motion ($\mu_{\delta}$ = 27.7~mas/yr) which differs very strongly
from the field. It turns out that Ruprecht 147 is an old cluster ($\log t$
= 9.39), at 175~pc from the Sun with an angular diameter of about
2.5~degrees i.e., 3 times larger than in Lyng\aa{}~(\cite{lyn87}).

\begin{table}

\caption{\label{tbl:rad}Radii (in pc) of clusters ($R_{cl}$) and associations
($R_{ass}$). Standard deviations are given in brackets.}

\begin{tabular}{lccc}
\hline
&
\multicolumn{2}{c}{New}&
 Lyng\aa{}\tabularnewline
\cline{2-3}
&
 Core &
 Corona &
 (1987)\tabularnewline
\hline
$\overline{R_{cl}}$ (10 smallest) &
 0.5 (0.1) &
 0.9 (0.2) &
 0.5 (0.3)\tabularnewline
$\overline{R_{cl}}$ (10 largest) &
 5.8 (2.4) &
 16.8 (2.4) &
 6.1 (4.9)\tabularnewline
$\overline{R_{cl}}$ ($d<450pc,N=67$)&
 1.6 (1.3) &
 5.1 (4.3) &
 3.6 (3.8)\tabularnewline
$\overline{R_{cl}}$ ($all,N=510$) &
 2.0 (1.3) &
 5.0 (3.2) &
 2.7 (2.7)\tabularnewline
$\overline{R_{ass}}$ ($all,N=9$) &
 6.9 (2.3) &
 33.2 (21.7)&
 --  \tabularnewline
\hline
\end{tabular}
\end{table}

\subsection{Statistic properties of the cluster sample\label{sec:biases}}

In order to find statistical relations involving sizes of open clusters,
we excluded 9 objects from our original sample since they are generally accepted
to be associations. Further, the two closest clusters, the Hyades
and Collinder~285 (the UMa cluster), are missing in our list. Since
they occupy large areas on the sky, a specific technique of membership
determination is required for them. Although the parameters needed
can be obtained from published data, we prefer, from the point of
view of data homogeneity, to not include these clusters
in the sample. So, the total cluster sample contains 641 open clusters.

Compared to previous work, this is the largest sample of galactic
open clusters ever used to find out statistical correlations including
cluster sizes. Studying the properties of open clusters, Lyng\aa{}~(\cite{lyn82})
and Janes et al.~(\cite{janea88}) were limited to samples of about
400 objects, mainly due to a lack of ages and/or distances for clusters.
The sample of Tadross et al.~(\cite{tad02}) includes 160 clusters
for which $UBV$ CCD observations are available. Although some of
the cluster parameters were redetermined by the authors, the samples
are compilations of published data and therefore, they are neither
complete nor homogeneous. From this point of view, a reliable statistical
proof of apparent correlations is rather difficult.

It is obvious, that the quantitative expressions derived below in
\S~\ref{sec:location} and \S~\ref{sec:age} depend strongly on
the conventions used for the definition of cluster sizes. Thanks
to the homogeneity of the data, it is expected that significant
correlations, derived with these data, indicate real trends. Nevertheless,
studying the distribution of cluster sizes, we must take into account
biases due to the relatively bright completeness limit of the \ascc at
about $V=11.5$. These biases occur in any kind of magnitude-limited
surveys, but in our case they become significant at relatively small
distances. The first one, an apparent {}``rejuvenation'' of the
cluster sample with increasing distance from the Sun, is illustrated
in Fig.~\ref{fig:lgt-dmod} and discussed in \S~\ref{sec:compl}.

\begin{figure}
%\selectlanguage{russian}
\includegraphics[%
  bb=172bp 53bp 561bp 476bp,
  clip,
  width=100cm,
  height=0.95\linewidth,
  keepaspectratio,
  angle=270]{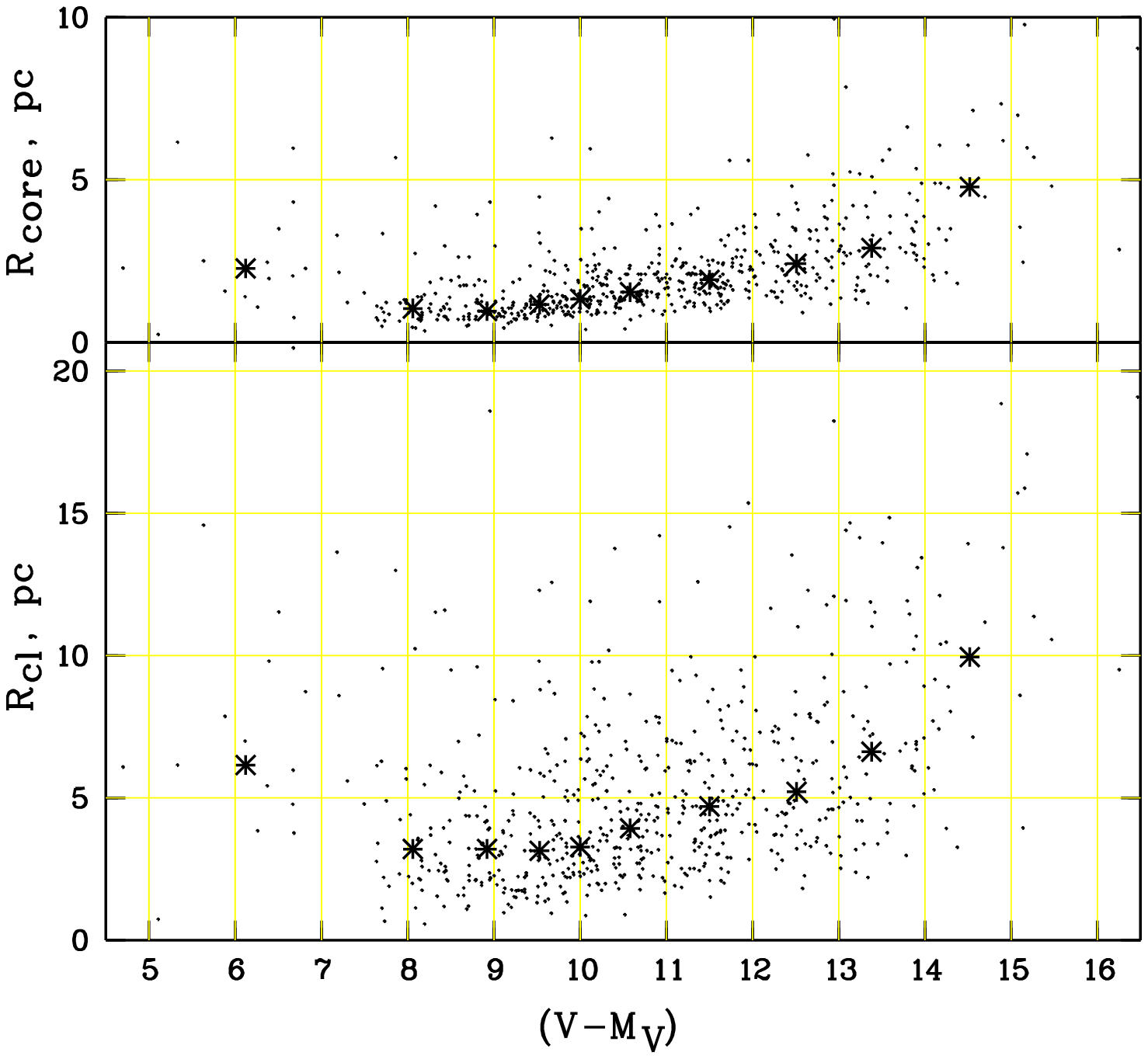}

\selectlanguage{english}
\caption{\label{fig:ralin}Linear radius of open clusters (bottom) and of
their cores (top) versus distance modulus. Dots are individual clusters,
asterisks mark the corresponding medians in bins of distance moduli.}
\end{figure}

Second, due to the low density of bright stars (about 150 stars/sq.deg
in the Galactic plane, H{\o}g et al.~\cite{tyc2}), we could not resolve open clusters with an
angular radius smaller than 0.08 degrees: one of our criteria for
the detection of a cluster was the presence of at least three $1\sigma$-members
in already known clusters and at least 8 members in newly detected
clusters. This is the reason for the absence of small clusters at
large distances in Fig.~\ref{fig:ralin}. Provided that the frequency
of small clusters at large distances is comparable to what we observe
within 400~pc from the Sun, there are about 10-15 clusters (i.e.,
4-6\%) still to be discovered within 850~pc. Of course, the number
of missing clusters grows rapidly with distance, and our sample becomes
more and more biased towards large young clusters (cf. Fig.~\ref{fig:ralin}).

The other bias in the determination of sizes arises due to expected
effect of mass segregation in open clusters. This means that at
larger distances we do not see faint members located in outer regions
of a cluster and thus we will systematically underestimate its size.
This bias is also a function of the limiting magnitude of the input
catalogue as well as of distance (or, more precisely, of the distance
and interstellar extinction, i.e. of distance modulus) of a cluster, and its influence
is difficult to estimate quantitatively. 

Finally, since linear sizes
of clusters are derived from angular sizes and distances, their accuracy
decreases with increasing distance. If not taken into account, this
obvious fact may lead to a misinterpretation of the apparent distribution
of linear sizes.

Assuming that these biases affect the size determination of all clusters
in a similar way, we can, however, study the distribution of linear
radii of different cluster groups provided that they have a comparable
distribution with respect to their distance moduli. In this case
the impact of biases onto the solution can be better taken into account.

\section{Relations between cluster size and the location in the Galaxy\label{sec:location}}

Throughout the paper we use the rectangular coordinate system $X,Y,Z$
with origin in the barycentre of the Solar system, and axes pointing
to the Galactic centre ($X$), to the direction of Galactic rotation
($Y$), and to the North Galactic pole ($Z$). Galactocentric distance
($R_{G}$) and distance from the symmetry plane ($|Z^{\prime}|$)
are computed for each open cluster under the assumption that the Sun
is located 8.5~kpc from the Galactic centre and 22~pc above the
symmetry plane of the cluster system (Paper IV).

\subsection{Cluster radius versus Galactocentric distance\label{sec:GC}}

The discussion on a possible dependence of the linear sizes of open
clusters from the Galactocentric distance has a long and controversial
history. Considering the sample of 150 open clusters from the Becker
\& Fenkart~(\cite{baf71}) catalogue, Burki \& Maeder ~(\cite{burki76})
concluded that the size of the youngest clusters ($\log t<7.20$)
increases with distance from the Galactic centre. Based on a sample
of about 400 clusters, Lyng\aa{}~(\cite{lyn82}) found that large
clusters ($R_{cl}>5$~pc) are mainly located outside the Solar orbit.
Using the same sample but with redetermined sizes and distances for
clusters, Janes et al.~(\cite{janea88}) did not reveal any significant
relation between cluster size and Galactocentric distance. With
a sample of 160 open clusters, Tadross et al.~(\cite{tad02}) found
a correlation between cluster size and Galactocentric distance for
clusters over the whole range of ages, whereas Nilakshi et al.~(\cite{nilak})
could confirm an increase of cluster sizes at $R_{G}>$ 9.5~kpc.

Since we are interested to find a possible gradient of linear cluster
sizes as a function of the Galactocentric distance, we exclude clusters
with $|Y|>$ 2~kpc. Although at large distances from the Sun, those
clusters may have a Galactocentric distance comparable to that of the
Sun. In this case they do not contribute effectively to the analysis
but introduce an additional noise due to uncertain parameters. Moreover,
considering a stripe along the Galactic radius, we are safer to assume
that the biases described in \S~\ref{sec:biases} are symmetrical
to the Sun's location, and possible differences in the sizes of clusters
for the inner ($R_{G}<$ 8.5 ~kpc) and outer ($R_{G}>$ 8.5~kpc)
disk should be real, if found in the first place.

\begin{figure}
%\selectlanguage{russian}
\includegraphics[%
  bb=55bp 51bp 566bp 584bp,
  clip,
  width=100cm,
  height=0.95\linewidth,
  keepaspectratio,
  angle=270]{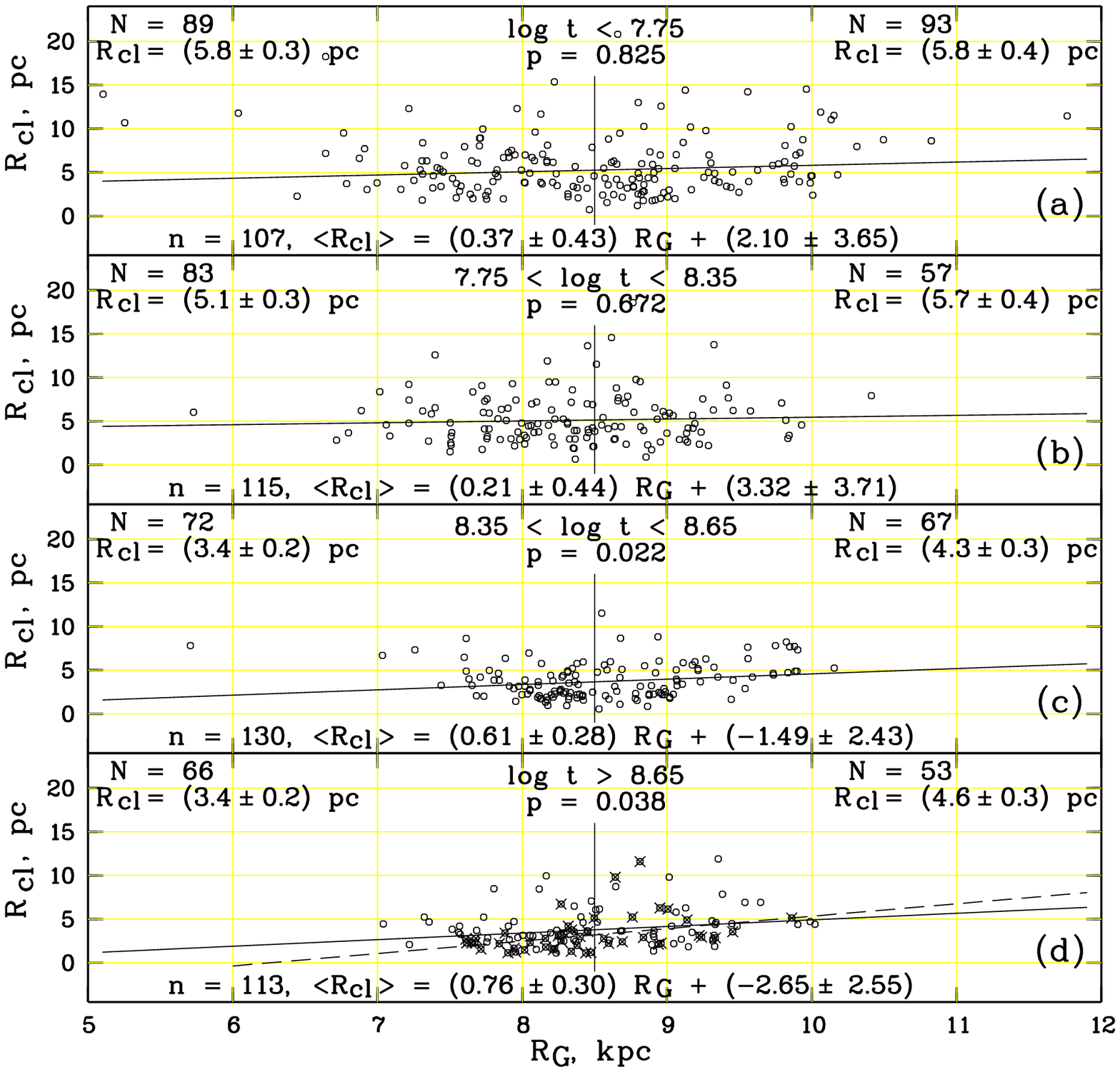}

\selectlanguage{english}
\caption{\label{fig:rl2gc} Linear radius of clusters versus Galactocentric
radius ($R_{G}$) for four age groups as indicated at the top of each
panel. Circles mark all clusters with $|Y|<$ 2~kpc. The line at
$R_{G}$ = 8.5~kpc divides clusters into {}``inner'' and {}``outer''
subsamples. For each subsample, $N$ gives the number of clusters,
$R_{cl}$ is the mean cluster radius. 
%whereas $\overline{|Z^{\prime}|}$ and $\sigma(Z^{\prime})$ are the
%mean distances from the symmetry plane and the dispersion of $Z^{\prime}$
%coordinates, respectively. 
$p$ is the probability that {}``inner''
and {}``outer'' clusters stem from the same statistical sample.
The solid lines are regression lines fitting the observed distributions
of the linear radii for clusters within $7 < (V-M_V) < 12$. The parameters
of the regression lines and the number of clusters included in the solution
are shown at the bottom of the corresponding panel.
For panel (d) crosses mark clusters older than $\log t=8.85$, and
the broken line is the corresponding regression line.}
\end{figure}

In order to compare the distributions of linear sizes of clusters
in the inner ($N$ = 310 clusters) and outer ($N$ = 270 clusters)
disk, a Kolmogorov-Smirnov ($K-S$) test was applied. We determined
a probability  $p$ = 0.002 for the null hypothesis that the linear
radii of {}``inner'' and {}``outer'' subsamples are drawn from
the same distribution. Further $K-S$ tests showed that the
differences in the distributions were caused by
clusters with age $\log t > 8.35$.
The relation between cluster radius
and Galactocentric distance is shown in Fig.~\ref{fig:rl2gc} for
four different age groups of clusters. The corresponding
least-square regression lines were calculated for clusters with
$7 < (V-M_V) < 12$. This region was chosen as a compromise: on one hand,
an $R_G$ spread had to be kept as large as possible since we looked for a
large scale effect; on the other hand, an impact of the biases
(see Fig.~\ref{fig:ralin}) must be minimised. Provided that the biases have a
similar influence on ``inner'' and ``outer'' subsamples, the compromise
is acceptable.

The age limits were not chosen arbitrarily. Analysing the kinematics
of open clusters in Paper IV, we derived a rotation velocity of the
cluster system of 234~km/s at the Galactocentric distance of the
Sun which corresponds to a rotation period $P_{GR0}$ of about 225
Myr around the Galactic centre. The youngest group ($\log t\leq$
7.75) includes clusters younger than 0.25$P_{GR0}$. The clusters
of the groups (b) and (c) have ages from 0.25$P_{GR0}$ to 1$P_{GR0}$
and from 1$P_{GR0}$ to 2$P_{GR0}$, respectively; whereas the cluster
in group (d) are older than 2$P_{GR0}$. Although the selection effects
and biases are clearly seen (no small clusters at larger distances
from the Sun, no old clusters at distances larger than 1.5~kpc from
the Sun), they affect the {}``inner'' and {}``outer'' cluster
subsamples in a similar way. Whereas the younger cluster groups do
not show any significant correlation between their size and Galactocentric distance,
a probable dependence appears after the first revolution around the
Galactic centre and becomes significant (at 2.5$\sigma$-level) when
clusters passed (survived) two revolutions. The old clusters in the
inner disk are on average smaller ($\overline{R_{cl}}=3.8\pm0.2$~pc)
than in the outer disk ($\overline{R_{cl}}=4.6\pm0.3$~pc), and the
probability that both cluster groups stem from the same statistical
sample is less than 4\%.

According to Friel~(\cite{friel}) no clusters older than $\log t=8.9$~Myr
(about the age of the Hyades) have been found within 7.5~kpc from
the Galactic centre. Our data support this result, though we cannot
exclude the possibility that some small old clusters could be located
at smaller $R_{G}$ and will be found with deeper surveys in the future.
They may not have been discovered yet, due to their small size and
low contrast to the field,
or due to clouds in the line of sight.
Extrapolating the relation between cluster
size and galactocentric radius derived for the old clusters (Fig.~\ref{fig:rl2gc},
panel (d)), we conclude that for old clusters ($\log t>8.65$, $\overline{\log t}=8.86$),
the limiting distance from the Galactic centre should be about 3.5~kpc.
By continously excluding younger clusters from the subsample (d),
we obtained steeper and steeper slopes of the relation and larger
limiting radii $R_{G}$. For $\log t>8.85$ ($\overline{\log t}=9.02$),
the limiting Galactocentric radius is about 6.3~kpc. Although the
number of clusters included in the latter case is relatively small ($N=45$),
the slope is significant ($\overline{R_{cl}}=(1.42\pm0.49)R_{G}-(8.90\pm4.16))$.
In other words, no clusters older than 1~Gyr should exist at galactocentric
radius less than 6~kpc.

\subsection{Cluster radius versus distance from the symmetry plane of the cluster
system\label{sec:zdist}}

\begin{figure}
%\selectlanguage{russian}
\includegraphics[%
  bb=55bp 51bp 566bp 584bp,
  clip,
  width=100cm,
  height=0.95\linewidth,
  keepaspectratio,
  angle=270]{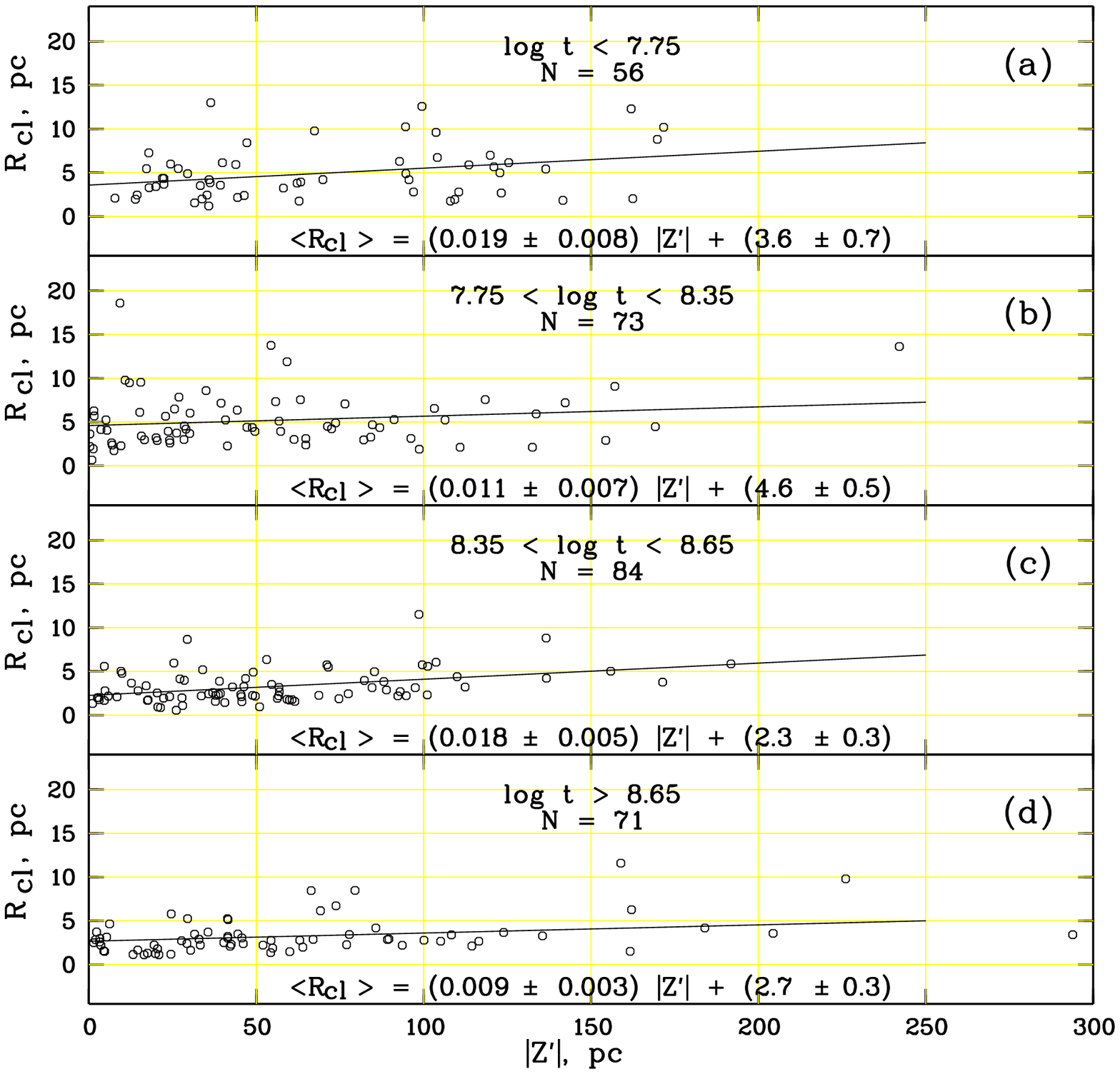}

\selectlanguage{english}
\caption{\label{fig:rl2z0}Linear radius of clusters versus distance from
the symmetry plane of the cluster system for four age groups as indicated
at the top of each panel. Circles mark all clusters with $7.0<(V-M_{V})<10.5$
. $N$ gives the number of clusters in each subsample, and the solid
line is the corresponding regression line fitting the observed distributions.
The parameters of the regression lines are given at the bottom of
each panel. Note on panel (d): the cluster NGC~2682 with $|Z^{\prime}|$
= 502~pc is not shown though it was included into the regression
calculation. }
\end{figure}

Considering the sizes of clusters as a function of the distance from
the Galactic plane, Janes et al.~(\cite{janea88}) found that small
clusters show a stronger concentration to the Galactic plane, independent
of age. A few large clusters are either old clusters at larger distances
from the Galactic plane or young clusters located close to the Galactic
plane. Tadross et al.~(\cite{tad02}) came to similar conclusions,
whereas no correlation between cluster sizes and $Z$ coordinates
was found by Nilakshi et al.~(\cite{nilak}). Lyng\aa{}~(\cite{lyn82})
presented a figure showing a distribution of cluster ages versus distance
from the Galactic plane, and he also distinguished between different cluster
sizes. No dependence between sizes and distances
from the Galactic plane is visible in this figure, and Lyng\aa{}~(\cite{lyn82})
did not comment this issue. We should note that large old clusters
from the samples of Janes et al.~(\cite{janea88}) and Tadross et al.~(\cite{tad02})
are located at such distances from the Sun (mainly from 1~kpc to 4~kpc)
where the cluster samples are highly biased by  incompleteness.
Hence, an absence of small clusters at large $Z$ distances could
be real or simply an apparent
trend due to selection effects and/or biases.

In order to minimise correlations due to different $R_{G}$ and the
biases described above, we consider clusters with $7.0<(V-M_{V})<10.5$,
i.e. in a $(V-M_{V})$ range where our sample is practically complete.
Although the mean cluster size $R_{cl}$ is computed as $4.1\pm0.2$~pc,
the clusters at $|Z^{\prime}|<50$~pc are, on average, smaller ($R_{cl}=3.6\pm0.2$~pc)
than those at $|Z^{\prime}|>100$~pc ($R_{cl}=5.2\pm0.4$~pc). The
distribution of cluster sizes versus distance from the symmetry plane
(where $Z^{\prime}$ = 0) is shown in Fig.~\ref{fig:rl2z0} for different
ages. According to Fig.~\ref{fig:rl2z0}, a systematic trend of cluster
sizes with increasing $|Z^{\prime}|$ can be observed for all clusters
in the Solar vicinity, and this correlation becomes highly significant for
clusters older than $\log t>8.35$, which already survived at least
one rotation around the Galactic centre.

\subsection{Cluster radius versus Galactocentric radius and distance from the
symmetry plane of the cluster system\label{sec:galrad}}

\begin{figure}
%\selectlanguage{russian}
\includegraphics[%
%  bb=66bp 42bp 564bp 583bp,
  bb=50bp 42bp 564bp 583bp,
  width=100cm,
  height=0.95\linewidth,
  keepaspectratio,
  angle=270]{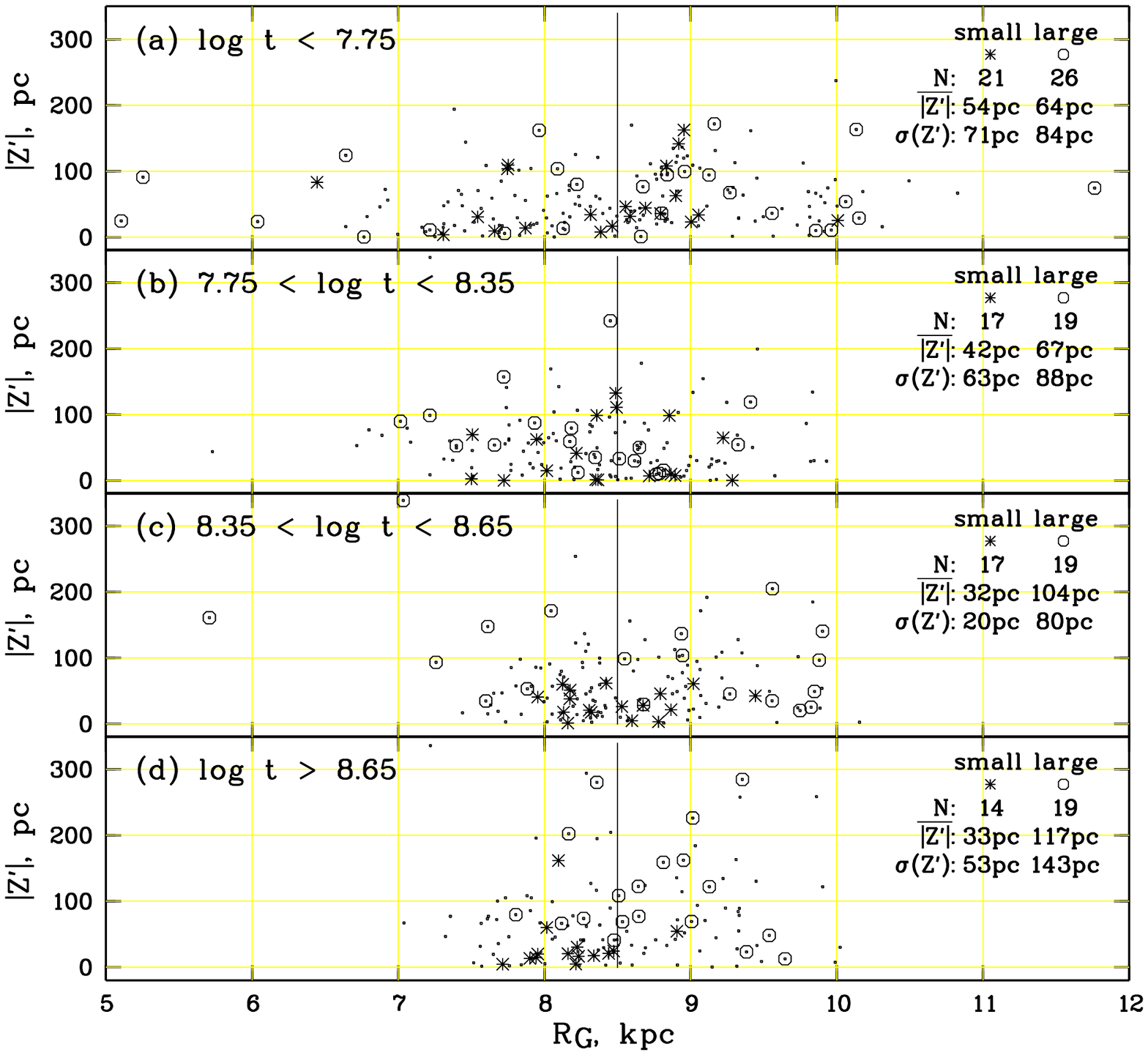}

\selectlanguage{english}
\caption{\label{fig:rg2z0}Distance of open clusters from the symmetry plane
versus Galactocentric radius for four age groups as indicated at the
top of each panel. Dots mark all clusters with $|Y|<$ 2~kpc. Circles
are {}``large'' clusters, asterisks mean {}``small'' clusters.
For the definition of {}``large'' and {}``small'' see text. In
each panel, $N$ is the number of {}``small'' or {}``large'' clusters,
$\overline{|Z^{\prime}|}$ is the mean distance from the symmetry
plane, and $\sigma(Z^{\prime})$ is the dispersion of $Z^{\prime}$.
The line at $R_{G}$ = 8.5~kpc divides clusters into {}``inner''
and {}``outer'' subsamples.}
\end{figure}

In order to check whether a multi-parameter correlation can be observed
for cluster sizes in our data set, we considered the same subsamples
as in \S~\ref{sec:GC}. The distribution of $|Z^{\prime}|$ distances
versus Galactocentric radius $R_{G}$ is shown in Fig.~\ref{fig:rg2z0}.
For each age group we computed the mean cluster radius and the corresponding
standard deviation. We call the clusters {}``large'' or {}``small''
if their linear radii differ from the corresponding mean by at least
one standard deviation. These clusters are indicated in Fig.~\ref{fig:rg2z0}
by different symbols. As expected, the youngest clusters (a) are more
strongly concentrated to the plane of the symmetry than the oldest
clusters (d). Small and large clusters of the youngest group show
a similar distribution with $R_{G}$ as well as with $|Z^{\prime}|$.
The location of young clusters is not uniform in the $R_{G}-|Z^{\prime}|$
space but shows links of the clusters to their formation places rather
than any systematics. In contrast, the spatial distributions of small
and large clusters differ in the oldest cluster group. The small old
clusters are more concentrated to the symmetry plane, whereas large
old clusters are found at large distances from this plane, though
the spread of $Z^{\prime}$ coordinates is large, too. Except in one
case, all large old clusters are at $R_{G}>8$~kpc. Such a distribution
would appear if the interstellar extinction were much stronger at
$7.5<R_{G}<8.0$ than at $9.0<R_{G}<9.5$. In this case, one would
expect to see this effect for other age groups, also. Since we do
not observe a similar evidence for younger clusters, we conclude that the
absence of large old clusters at smaller $R_{G}$ indicate a real
trend.

Summarising the findings of this section, we propose the following
scenario for cluster evolution within the Solar neighbourhood. Clusters
are formed within a thin disk inside as
well as outside the Solar orbit. Their initial sizes do not show significant
correlations with the $R_{G}$-- and $|Z^{\prime}|$- coordinates.
After one revolution around the Galactic centre, the distribution
of cluster sizes becomes more systematic with respect to the cluster
location: small clusters are more probably found at low $Z^{\prime}$
coordinates, whereas large clusters are generally located at larger
$Z^{\prime}$ and/or at larger $R_{G}$. Some of the large young clusters
probably dissolve during the first revolution around the
Galactic centre. However, they have a higher chance to survive encounters
with molecular clouds and the impact of Galactic tidal forces and
reach higher ages if their orbits are outside the Solar orbit and are inclined
to the Galactic plane (cf. Fig.~\ref{fig:rg2z0}). The concentration of
small clusters with $R_G < 8.5$~kpc to the symmetry plane supports
indirectly the conclusions drawn from the simulations by 
Spitzer \& Chevalier~(\cite{spch73}) that compact clusters survive
against external shocks for a longer time.

\section{Relations between cluster size and age\label{sec:age}}

\begin{figure}
%\selectlanguage{russian}
\includegraphics[%
  bb=180bp 52bp 558bp 608bp,
  clip,
  height=0.95\linewidth,
  keepaspectratio,
  angle=270]{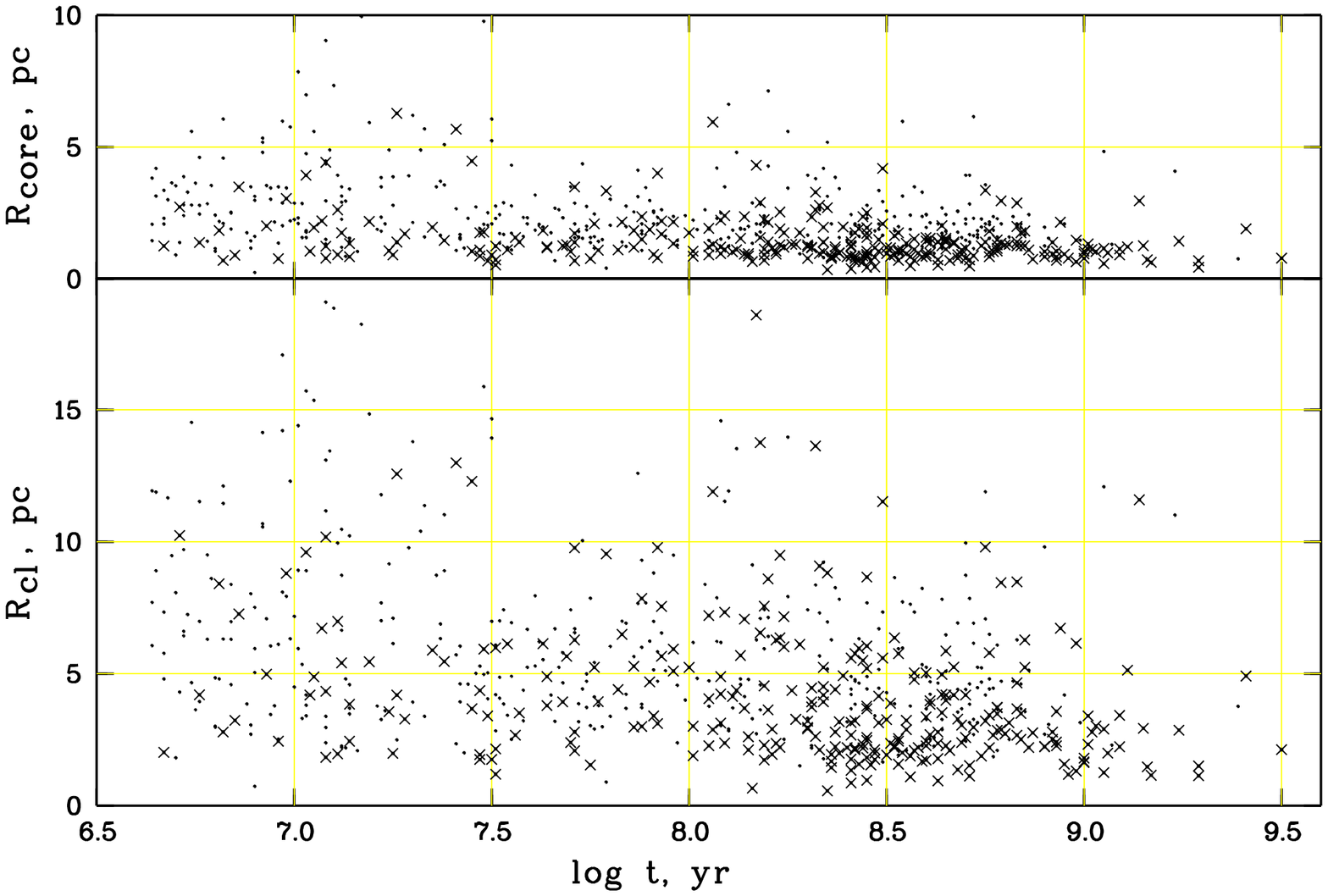}

\selectlanguage{english}
\caption{\label{fig:rllgt}Linear radius of open clusters (bottom) and of
their cores (top) versus cluster ages. Crosses mark clusters with
$7.0\leq(V-M_{V})\leq10.5$.}
\end{figure}

Neither Lyng\aa{}~(\cite{lyn82}), Janes et al.~(\cite{janea88}),
Tadross et al.~(\cite{tad02}) nor Nilakshi et al.~(\cite{nilak})
found a correlation between cluster sizes and ages. This is not very
surprising since, as discussed in the previous section, cluster sizes
seem to show a multi-parametric dependence. The separation of different
effects is rather difficult, especially, if one cannot rely on a complete
and unbiased sample as well as on the homogeneity of cluster parameters.
Nevertheless, the relations between the linear radii and the location
of open clusters in the Galaxy indicate a correlation with cluster
age. This gives us a strong hint to look more carefully whether any
direct dependence of cluster sizes on age can be found in our data.
Since the age for each open cluster of our sample was determined by
the same method and linear sizes $R_{cl}$, $R_{core}$ were derived
via of the same conventions, the data provide the best preconditions
for checking whether real trends in structural parameters exist, depending
on ages.

\begin{figure}
%\selectlanguage{russian}
\includegraphics[%
  bb=62bp 93bp 568bp 758bp,
  clip,
  height=0.95\linewidth,
  keepaspectratio,
  angle=270]{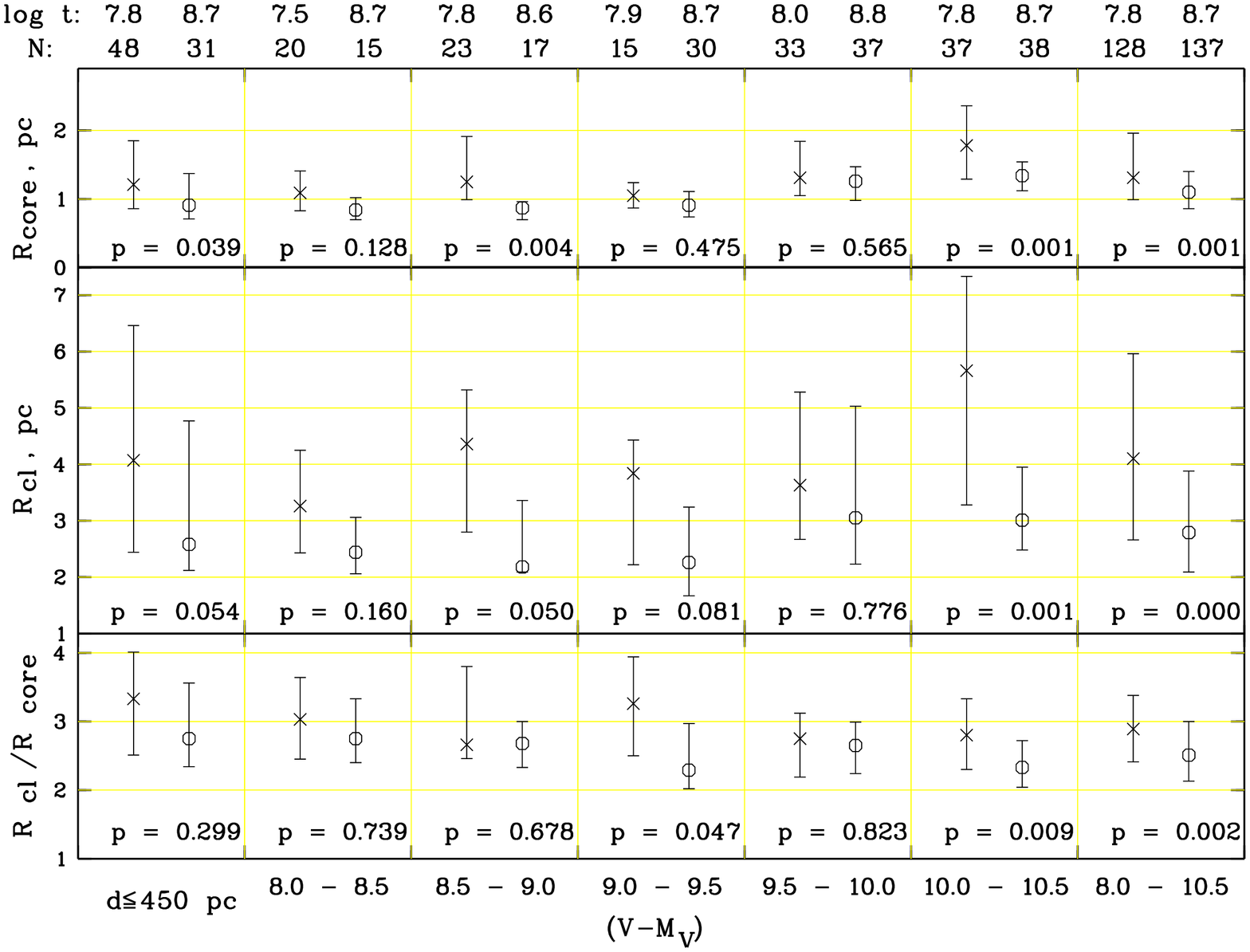}

\selectlanguage{english}
\caption{\label{fig:radlgt}Structural parameters of open clusters for two
age groups: a young group with $\log t\leq8.4$, and an old group
with $\log t>8.4$. The upper panel is for cluster core radii $R_{core}$,
the central - for linear radii of clusters $R_{cl}$, and the lower
- for ratios $R_{cl}/R_{core}$. Crosses and circles show the median
for young and old groups, respectively, and bars are $Q_{1}$, $Q_{3}$
quartiles computed for each age group and for different $(V-M_{V})$
bins (as indicated at the bottom). $\log t$ and $N$ at the top give
the average age and the number of clusters in the corresponding subsample,
and $p$ is the probability that the parameters of young and old groups
at a given $(V-M_{V})$ are drawn from the same distribution.}
\end{figure}

The distribution of clusters as a function of age is shown in Fig.~\ref{fig:rllgt}.
At the first glance, a decrease of cluster sizes and their scattering
with age may be suggested, though, the majority of large-radii clusters
are distant young ones. In order to minimise possible biases described
above, we selected clusters within 5 small bins of distance moduli
$\Delta(V-M_{V})=0.5$ and considered the linear radii of clusters
in two subgroups, younger and older than 250~Myr ($\log t=8.4$).
This age limit was chosen arbitrarily, only to provide more or less
comparable numbers of clusters in young and old groups. Additionally,
we considered all clusters within the range $(V-M_{V})$ = 8...10.5
as well as the nearby clusters with distances up to 450 pc from the
Sun. Since the cluster radii are not distributed normally but show
rather skewed distributions with long tails towards large radii, the
mean values of the subsamples are affected by extreme values of {}``outliers''.
The impact can be essential since the number of clusters in each subsample
is relatively small. Therefore, we considered the median, the first
$Q_{1}$ and third $Q_{3}$ quartiles of each data set. The results
are given in Fig.~\ref{fig:radlgt}. Applying the $K-S$ test, we
also computed the probabilities $p$ for the null hypothesis that
the linear radii of young and old groups at a given $(V-M_{V})$ are
drawn from the same distribution.

Although the cluster sizes $R_{cl}$ of two age groups do not always
differ significantly (i.e., $p>0.05$), the general tendency remains
remarkably constant: the younger groups have in average larger sizes
by a factor of $\approx1.6$ (1.2...2.0), and they show a larger spread
of sizes. This is also valid for the cluster cores, though, the effect
is smaller. The ratios $R_{cl}/R_{core}$ are more affected by poor
statistics, though they also indicate a similar tendency: their medians
range within 2.7...3.3 for the younger groups, and within 2.3...2.7
for the older clusters. We conclude that the results seem to indicate
trends in structural parameters depending on cluster ages.

\begin{figure}
%\selectlanguage{russian}
\includegraphics[%
  bb=65bp 91bp 555bp 678bp,
  clip,
  height=0.95\linewidth,
  keepaspectratio,
  angle=270]{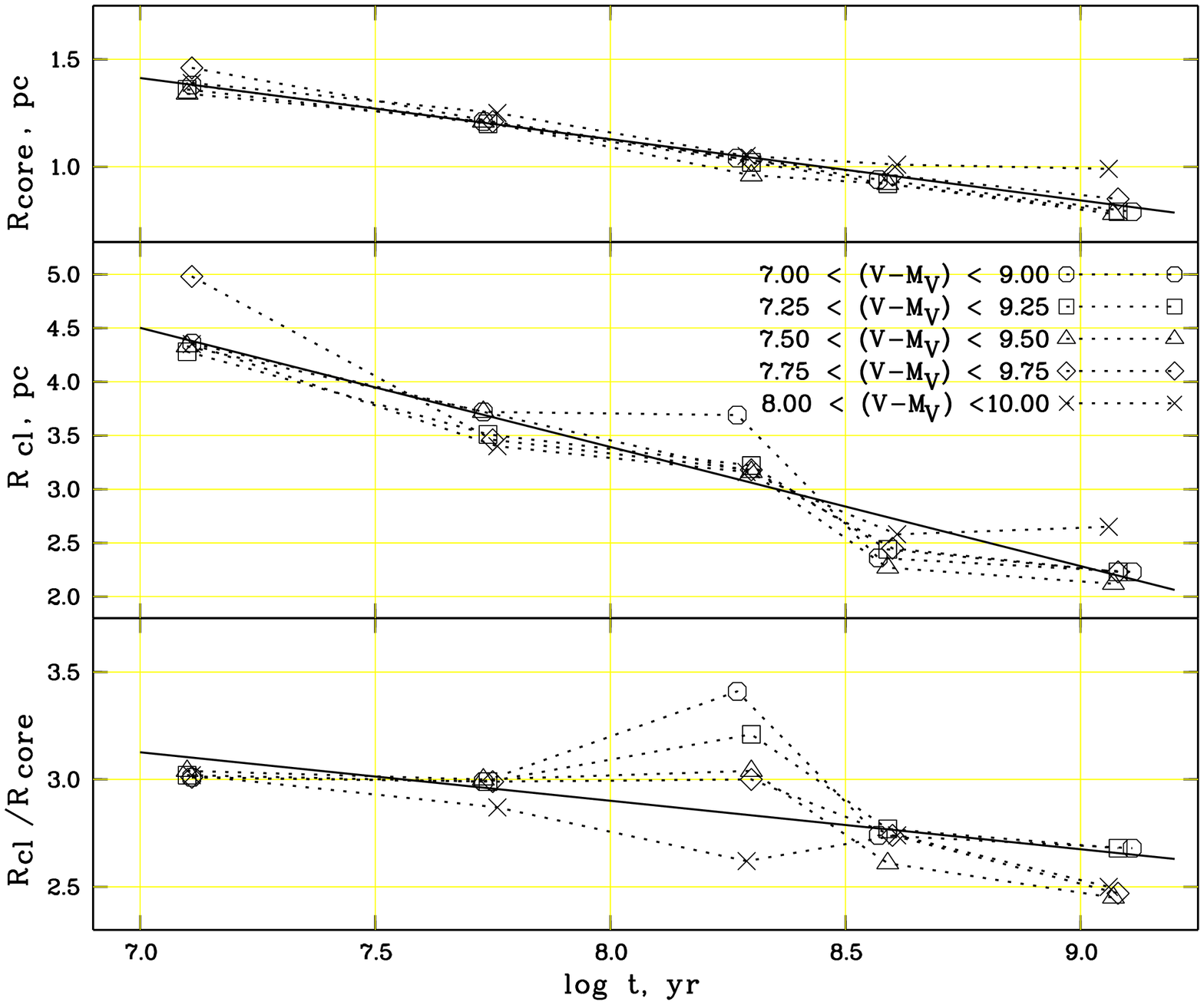}

\selectlanguage{english}
\caption{\label{fig:medradlgt}Structural parameters of open clusters versus
age. The upper panel is for linear radii of cluster cores $R_{core}$,
the central - for linear radii of clusters $R_{cl}$, and the lower
- for the ratios $R_{cl}/R_{core}$. Thin dashed lines connect the
medians determined for different age groups within one $(V-M_{V})$--bin.
The solid lines are the regression lines fitting the observed distributions
of the medians.}
\end{figure}

In order to obtain a quantitative expression for the dependence of
cluster sizes on the age and to check its significance, the clusters
were divided into 5 age groups with $\log t$ (i.e., $\leq7.50$,
7.50...8.00, 8.00...8.45, 8.45...8.80, $>8.80$). Again, these ranges
were chosen arbitrarily, as a compromise between the number of age groups
and numbers of clusters in each age group. Further, we selected clusters
with $(V-M_{V})$ from 7~mag to 10~mag binned in 5 overlapping groups
(see Fig.~\ref{fig:medradlgt}).
In each $age/(V-M_{V})_{i}$ subsample,
the median is determined for $R_{core}$, $R_{cl}$, and $R_{cl}/R_{core}$.
The corresponding $5\times5$ realisations of the medians are shown
in Fig.~\ref{fig:medradlgt}. The resulting relations between structural
parameters of open clusters and their ages $\log t$ can be approximated
by the following equations:

\begin{eqnarray}
R_{cl} & = & (-1.11\pm0.08)\,\log t+(12.26\pm0.67)\nonumber \\
R_{core} & = & (-0.28\pm0.02)\,\log t+(3.40\pm0.12)\label{eqn:msizeage}\\
R_{cl}/R_{core} & = & (-0.23\pm0.06)\,\log t+(4.70\pm0.44)\nonumber \end{eqnarray}
where $R$ is measured in parsec.
 For comparison, we recomputed the dependences by including all 209
clusters within $(V-M_{V})$ from 7~mag to 10~mag: \begin{eqnarray}
R_{cl} & = & (-1.26\pm0.30)\,\log t+(14.35\pm2.49)\nonumber \\
R_{core} & = & (-0.39\pm0.09)\,\log t+(4.54\pm0.78)\label{eqn:sizeage}\\
R_{cl}/R_{core} & = & (-0.26\pm0.10)\,\log t+(5.09\pm0.84)\nonumber \end{eqnarray}
 Due to {}``outliers'' (i.e., the clusters with large $R_{cl}$
and $R_{core}$), the regression lines are somewhat shifted but the
differences are still within the error range. Although all terms in
eqs.~(\ref{eqn:msizeage}), (\ref{eqn:sizeage}) are highly significant,
the derived relations are, of course, not universal. Based on other
definitions of $R_{cl}$ and $R_{core}$ or on a survey with different
completing magnitude, the relations may change. One may expect that
the impact would be stronger on the {}``zero-point''-term than on
the slopes, describing the correlation of radii with age which are, indeed,
the more important and interesting parameters. According to eqs.~(\ref{eqn:msizeage}),
(\ref{eqn:sizeage}), we conclude that, on average, the apparent linear
sizes of clusters and their cores are decreasing with time and that
the process is going faster for the cluster sizes themselves than for
their cores. The question is whether the averaged size of 
open clusters really become smaller
with age, or if this is an apparent trend due to e.g. mass segregation
effects acting differently in old and young clusters, or perhaps there
is a mixture of both effects.

\section{Mass segregation in open clusters\label{sec:masseg}}

\begin{figure}
%\selectlanguage{russian}
\includegraphics[%
  bb=134bp 70bp 540bp 555bp,
  clip,
  height=0.95\linewidth,
  keepaspectratio,
  angle=270]{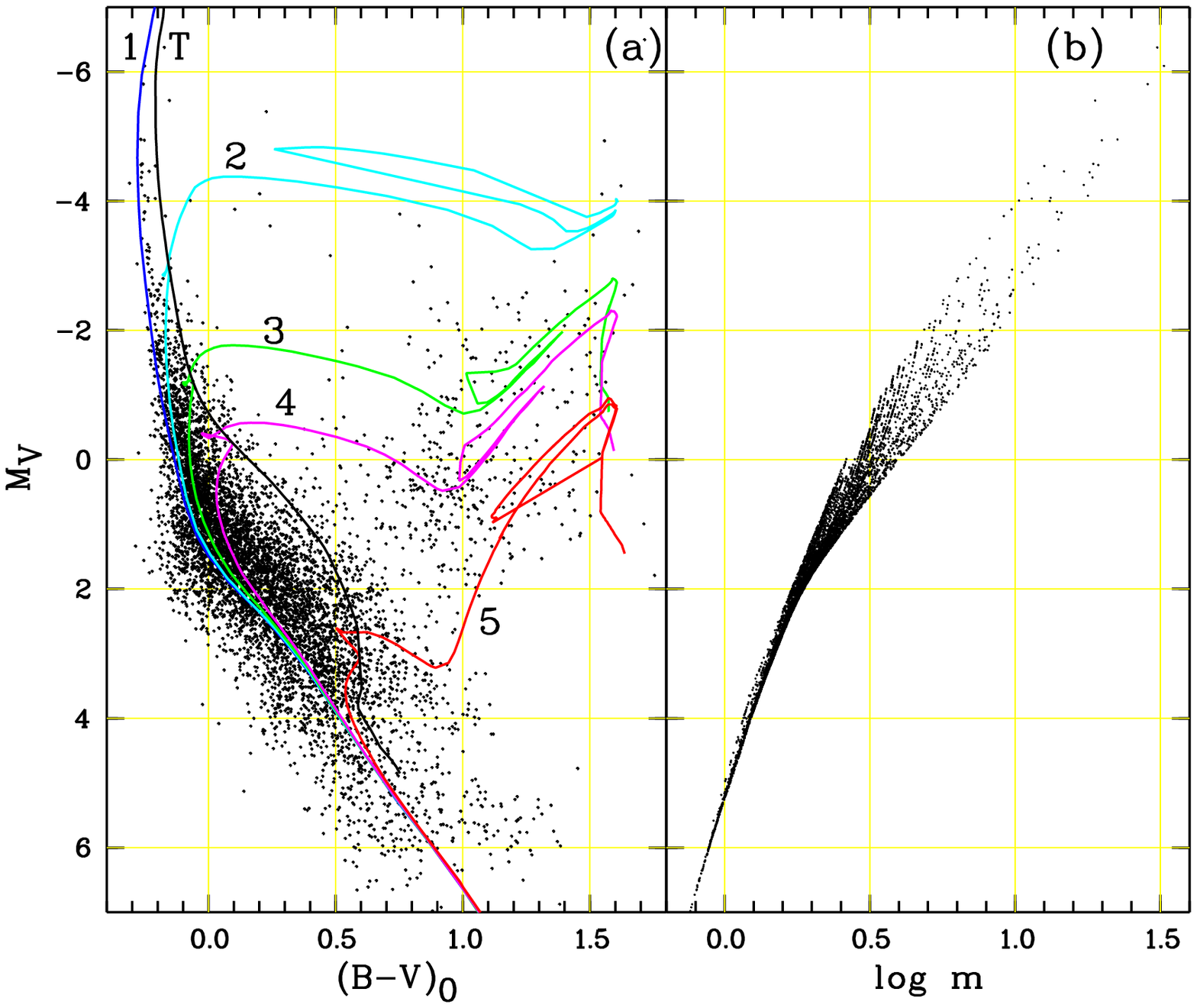}

\selectlanguage{english}
\caption{\label{fig:cmd-lgm}Absolute magnitude ($M_{V}$) and mass ($\log\, m$)
of the most probable members of clusters with $(V-M_{V})<10.5$. Panel
(a): CMD. Curves 1, 2, 3, 4, and 5 show the Padova isochrones for
$\log t$= 6.65, 7.75, 8.35, 8.65, and 9.50, respectively; the TAMS
is marked by T. Panel (b): Relation between absolute magnitude and
$\log\, m$ of the most probable MS members. $m$ is in units of solar
masses.}
\end{figure}

In order to quantify the effect of mass segregation in open clusters,
different approaches are usually applied. The majority of methods
is based on the comparison of the integrated profiles of the surface
density for stars with different mass, and on an analysis of the differences
in their concentration to the cluster centre (e.g., Mathieu~\cite{math84},
Sagar et al.~\cite{sagar88}, Raboud \& Mermilliod~\cite{ramer98a,ramer98b},
Hillenbrand \& Hartmann~\cite{hiha}, Bonatto \& Bica~\cite{bonb}).
Another approach considers
luminosity and/or mass functions and compares their slopes for cluster
stars located in the central and outer areas of clusters (Fischer
et al.~\cite{fishea98}, de Grijs at al.~\cite{deg02a,deg02b,deg02c}).
This method is indirect and rather difficult to apply, since a survey
of identical completeness is required for the central and outer regions.
Usually, this requirement is hardly achievable due to crowding effects
in the cluster center. Some authors (Sagar et al.~\cite{sagar88},
Hillenbrand \& Hartmann~\cite{hiha}, Slesnick et al.~\cite{shm02})
consider radial trends in the average stellar mass, though this method
needs not only a complete but also a deep survey due to the weak
significance of the effect.

Due to the relatively bright limiting magnitude of the \ascc, and
therefore, the relatively low average number of the most probable members
%(number of the $1-\sigma$-members ranges from 3 to 178, with an average of 17),
we cannot apply the methods described above to the majority of clusters
of our sample. Therefore, we need to introduce another parameter which
takes into account properties of our data. Then, having the uniform
data set of linear sizes and the information on spatial distribution
of the most probable members in each cluster, we can analyse the effects
of mass segregation and study general trends in the mass distribution
of members in open clusters of different ages.

For an easier interpretation of the results on mass segregation, we
computed masses of cluster members within a range of absolute magnitudes
which is typical for our sample. We used the Padova grid of overshooting
isochrones (Girardi et al. \cite{pad02}) with input parameters $m=0.15...66\, m_{\odot}$,
$Z=0.019$, $Y=0.273$. As we consider open clusters in a relatively
small range of distances from the Sun, a possible impact of metallicity
variation on mass determination can be ignored. In order to avoid the
strong uncertainty due to models for red giants, we did not include
members to the right of the TAMS. In Fig.\ref{fig:cmd-lgm}a we show
a combined CMD of the most probable members of open clusters with
$(V-M_{V})<10.5$ and a number of isochrones covering the complete
range of cluster ages in our sample. Fig.\ref{fig:cmd-lgm}b gives
the relation between absolute magnitude and stellar mass for cluster
members located to the left of the TAMS in Fig.\ref{fig:cmd-lgm}a.
Consequently, the main sequence (MS) members of the clusters of our
sample cover ranges $M_{V}\approx-6...+6$ mag and $\log m\approx1.5...-0.1$,
where $m$ is in units of solar masses.

\begin{figure}
%\selectlanguage{russian}
\includegraphics[%
  bb=84bp 52bp 548bp 675bp,
  clip,
  width=100cm,
  height=0.95\linewidth,
  keepaspectratio,
  angle=270]{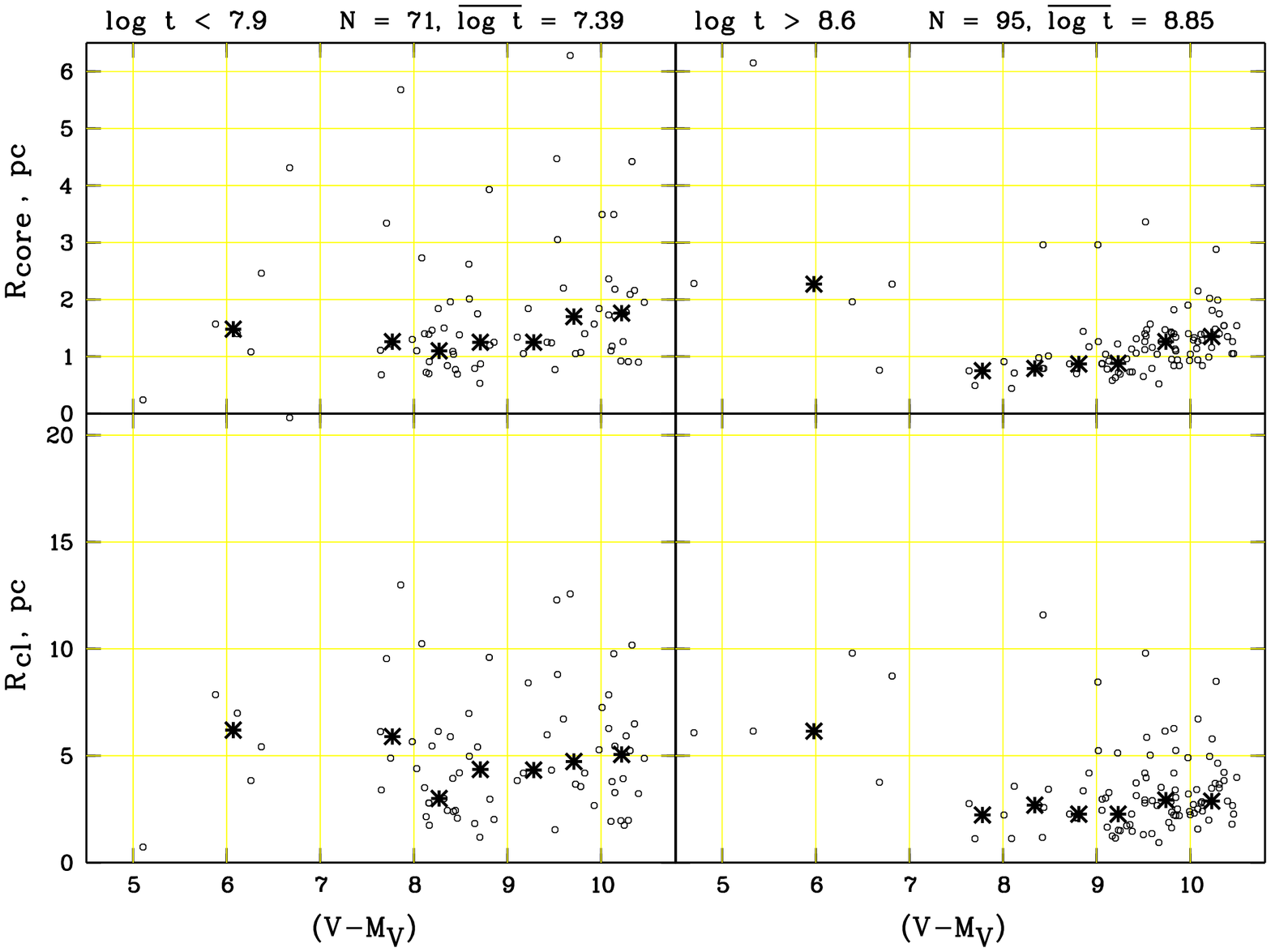}

\selectlanguage{english}
\caption{\label{fig:medrad2lgt}Linear radii of clusters (bottom panels) and
cores (upper panels) versus distance moduli for two different age
groups as indicated at the top. Circles are individual clusters, asterisks
show the corresponding medians of cluster and core radii. $N$ and
$\overline{\log t}$ are the numbers of clusters in each group and their
mean age.}
\end{figure}

Now we come back to the question at the end of the previous section.
Considering only clusters within the completeness area, let us compare
the correlations of sizes of the youngest and oldest clusters with
their distance from the Sun. If one assumes a higher concentration
of relatively massive stars (observable in the \ascc up to large
distances) to the cluster centre and a widely spread distribution
of fainter stars (missing in the \ascc at large distances), the linear
sizes of open clusters should decrease with increasing $(V-M_{V})$.
If the sizes of the oldest clusters would decrease faster than the
sizes of the youngest group, it would hint to a stronger mass segregation
in older clusters. The distribution of cluster sizes is shown in Fig.~\ref{fig:medrad2lgt}
for young ($\log t<$ 7.9) and old ($\log t>$ 8.6) clusters with
distance moduli $(V-M_{V}) < 10.5$. Unfortunately, due
to the relatively low spatial density of clusters, we are, in practice,
limited to a $(V-M_{V})$ range between 7.5 and 10.5 mag which, taking
into account the large scattering of cluster sizes, is rather small
to get a clear quantitative proof. Nevertheless, qualitative conclusions
seem possible.

\begin{figure*}
%\selectlanguage{russian}
\includegraphics[%
  bb=63bp 36bp 520bp 731bp,
  clip,
  height=0.95\linewidth,
  keepaspectratio,
  angle=270]{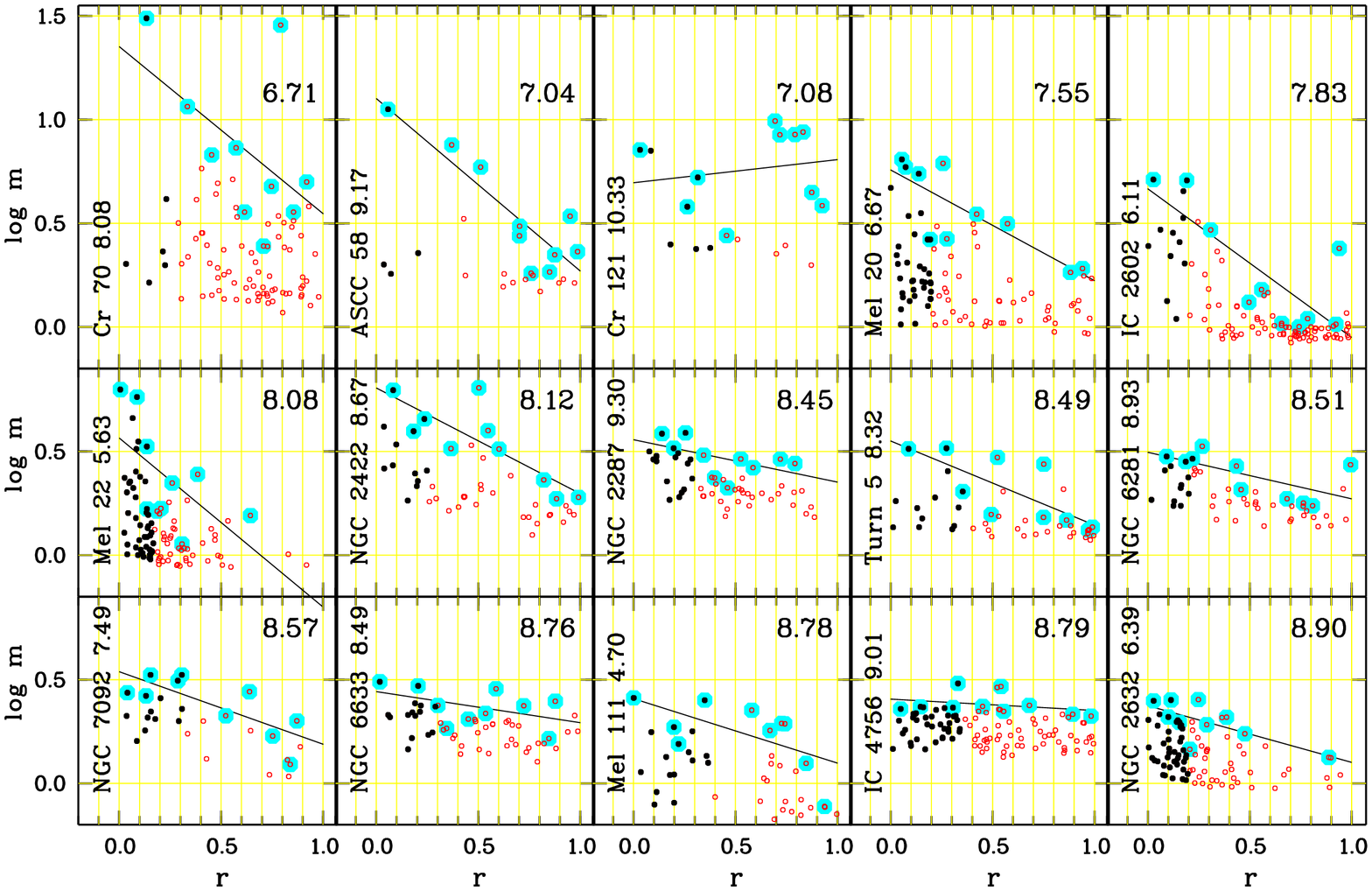}

\selectlanguage{english}
\caption{\label{fig:rr-lgm-p}The mass of the most probable members versus radial
distance from the cluster centre (in units of $R_{cl}$) for selected
clusters of different ages. In each panel, the cluster name and distance
modulus are shown on the left, whereas the cluster age is given on
top. Black dots indicate cluster members projected onto the core area,
(red) open circles mark members in the coronal area. Grey (cyan) circles
show members included in the calculation of the corresponding regression
shown as the straight line (see text for further explanation). Clusters
are sorted with increasing age.}
\end{figure*}

With the given completeness limit of the \ascc at about 11.5 mag,
the faintest stars, which we observe in a cluster at $(V-M_{V})$
= 6, are about $M_{V}$ = 5.5 corresponding to $m\approx0.9\, m_{\odot}$.
On the other hand, at $(V-M_{V})$ = 10.5 the observed sizes of clusters
are defined by stars brighter than $M_{V}$ = 1 with masses $m>2.5\, m_{\odot}$.
According to the median of radii, a few nearby clusters ($\overline{V-M_{V}}\approx6$)
in Fig.\ref{fig:medrad2lgt} (bottom panels) have, on average, comparable
sizes independent of their age. This can occur if stars with masses
$m\approx0.9\, m_{\odot}$ are observed at distances from the cluster
centre which are similar for relatively young ($\overline{age}\approx$
30 Myr) and old ($\overline{age}\approx$ 800 Myr) clusters. The situation
changes if we consider stars of larger masses which are at the magnitude limit
of the \ascc at distance moduli between 8 and 10.5. In older
clusters, the stars with masses between $1.3\, m_{\odot}$ ($M_{V}$
= 3.5, $(V-M_{V})=8$) and $2.6\, m_{\odot}$ ($M_{V}$ = 1, $(V-M_{V})=10.5$)
are concentrated to the cluster center much stronger (by a factor
of two) than stars of $0.9\, m_{\odot}$ at $(V-M_{V})=6$. On the
other hand, this effect is less significant in young clusters: the
cluster sizes defined by stars with masses 1.3...$2.6\, m_{\odot}$
are only slightly smaller than the sizes defined by stars with masses
$0.9\, m_{\odot}$. On average, young clusters are 1.6 times larger
than old clusters if we consider members with masses 1.3...$2.6\, m_{\odot}$
at $(V-M_{V})$ = 8...10.5. Although the scattering of radii is
rather high, especially for young clusters, this seems to be a general
trend.

For the cluster cores (Fig.~\ref{fig:medrad2lgt}, upper panels),
the trends are similar. An increase of core radii at $(V-M_{V})>9.5$,
probably comes from the definition we adopted for the cluster core.

We conclude that the apparent decrease of cluster sizes with increasing
$\log t$ observed in Fig.\ref{fig:medradlgt} and described by eqs.~(\ref{eqn:msizeage}),
(\ref{eqn:sizeage}) is generally caused by different concentration
of cluster members with masses 1.3...$2.6\, m_{\odot}$
with respect to the cluster centre: the concentration increases with
increasing age of clusters. On the other hand, for stars of about $0.9\, m_{\odot}$
the apparent distribution of linear radii does not differ considerably for young and old 
clusters in the Solar neighbourhood ($V-M_V < 7.5$). 
Therefore, the observed dependence of cluster sizes on age can be
explained by a stronger mass segregation of stars with $m > 1.3\, m_{\odot}$
in old clusters rather than by a decrease of the real sizes of clusters. 
Of course, we cannot exclude that,
with a deeper input catalogue and with a larger portion of very old clusters 
in a sample, a real decrease of the average cluster size with age can be 
found. This could be so, especially, if the location of clusters
in the Galaxy is taken into acount: the corresponding hints that large clusters
have a lower chance to survive tidal effects are obtained in \S~\ref{sec:location}.

More detailed conclusions on mass segregation at $m > 1.3\, m_{\odot}$
can be drawn if one considers the radial
distribution of the most massive stars in a cluster. To illustrate this approach
we show 15 clusters of different age in Fig.~\ref{fig:rr-lgm-p}. The clusters
have an extended magnitude range $\Delta V$>3 mag and
are presented in a sequence of increasing age. Although
the observed lower mass limit depends on the distance modulus, this
is of lesser importance for the following analysis since we are interested
in the upper part of the profiles (i.e., the most massive stars at
a given distance from the cluster centre) only.

Guided by Fig.~\ref{fig:rr-lgm-p} we use the slope of the {}``maximum
stellar mass -- distance from cluster centre'' relation
as a statistical parameter to quantify the mass segregation effect.
In order to compute this parameter, we subdivided the area of each cluster
into 10 concentric rings of variable width but containing an equal number
of the most probable members. This binning provides an unbiased sampling
both with respect to the variation of the density profile and to the representativity of
the mass distribution in a cluster. In each ring, the star
with maximum mass $\log\, m_{max}(r)$ was selected.
Here $r$ is the distance from the cluster centre in units of the
cluster radius i.e., $r=R/R_{cl}$, $0\leq r\leq1$. Of course,
the most reliable results can be expected for clusters with a sufficiently
large number of members covering an extended range of masses. As a
compromise, we included in our analysis only clusters having a main sequence 
extend larger than 3 mag in the \ascc (typically, 3.5 mag for $\log t>8$,
and 6 mag for younger clusters). 
In order to be certain that we
consider comparable mass ranges, we also required that at least one
cluster member must be less massive than $2\, m_{\odot}$. In total,
167 clusters meet these requirements.

For each cluster, we solved a system of linear equations describing
the variations of $m_{max}(r)$ as a function of the distance from the
cluster centre \begin{equation}
\log m_{max}(r)=b\times r+a\,\label{eqn:lgm-rr}\end{equation}
 with  parameters $a$ and $b$, where $b$ describes the
radial mass gradient $d\log m_{max}/dr$. For illustration, the relations
together with the cluster members included in the solution are shown
in Fig.~\ref{fig:rr-lgm-p}.

\begin{figure}
\includegraphics[%
  bb=50bp 65bp 565bp 363bp,
  clip,
  height=0.95\linewidth,
  keepaspectratio,
  angle=270]{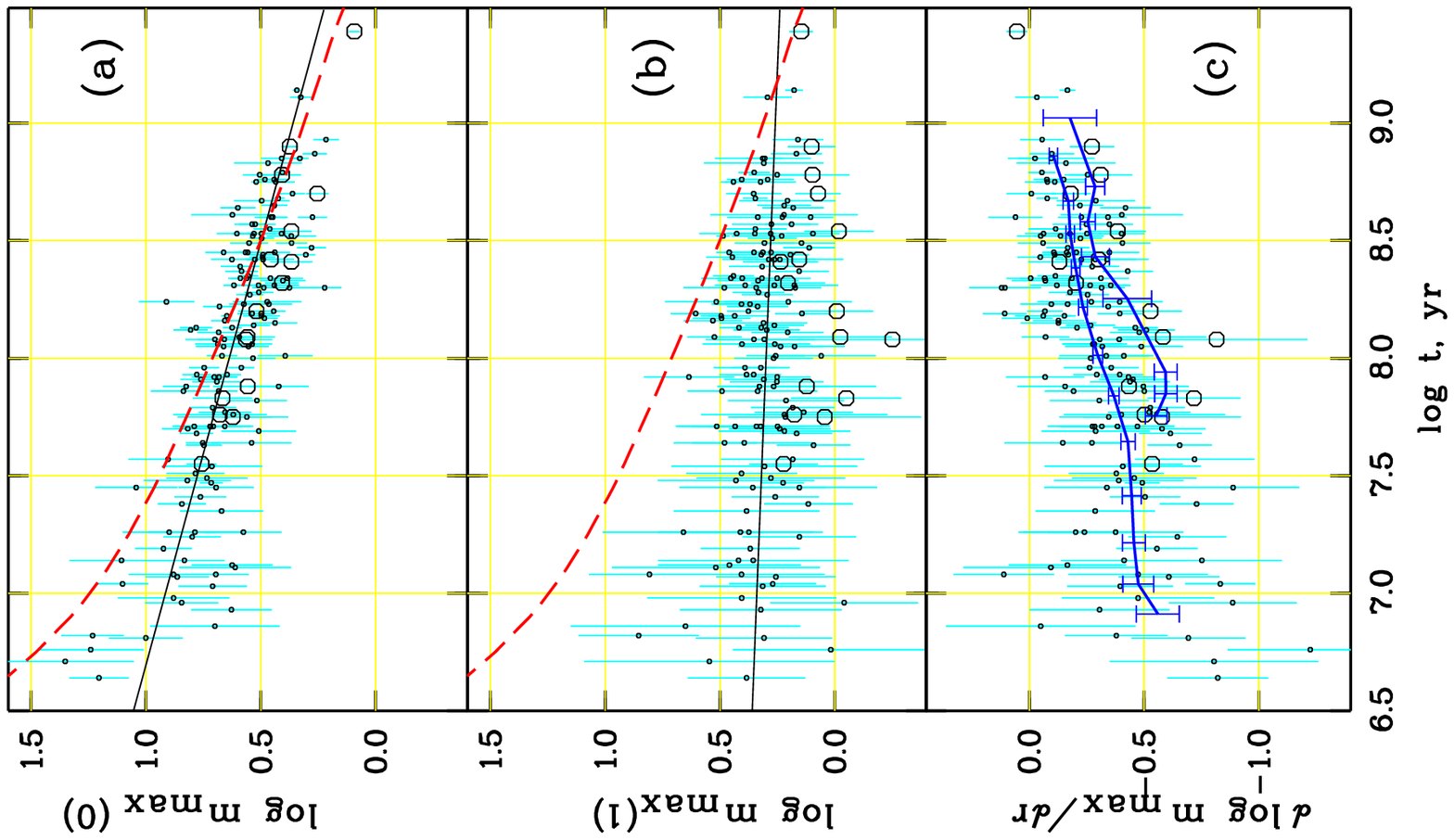}

\caption{\label{fig:lgt-lgm} The expected mass of the brightest stars
computed for the cluster centres (a) and edges (b) versus cluster age.
Panel (c) shows the variation of radial mass gradient $d\log m_{max}/dr$
with cluster age. Circles indicate all clusters included in the analysis,
large open circles mark the subsample of nearby
clusters ($(V-M_{V}) < $ 7.3) with the most extended Main Sequences.
The $rms$ errors of the individual data points are shown as 
light grey (cyan) bars. In panels (a) and (b): the dashed curves 
show the ``mass--MS-lifetime relation'' based on the Padova isochrones with 
overshooting for $Z=0.019$, solid lines present the solution of
equations(\ref{eqn:lgm-t0})-(\ref{eqn:lgm-t1}). In panel (c): 
the curves show the running average
of the radial mass gradient computed with a ($\log t$)-bin of
0.6 and a step of 0.2 for nearby clusters (lower curve) and
for all other clusters (upper curve).
The $rms$ errors of the corresponding averages
are shown with bars. 
}
\end{figure}

From the coefficients $a$ and $b$ of eq.~(\ref{eqn:lgm-rr}) we
computed the maximum mass of a cluster member expected in the cluster
centre ($r$ = 0) and at the cluster edge ($r$ = 1). The dependence of
these quantities
on cluster age can be well approximated by the equations

\begin{eqnarray}
\log m_{max}(0) & = & (-0.29\pm0.02)\,\log t_{6}+(1.20\pm0.04),\label{eqn:lgm-t0}\\
\log m_{max}(1) & = &(-0.04\pm0.02)\,\log t_{6}+(0.38\pm0.04)  \label{eqn:lgm-t1}
\end{eqnarray}
 where $t_{6}$ is the cluster age in units of Myr. The corresponding
solutions are shown in Fig.~\ref{fig:lgt-lgm}, in panel (a) for cluster
centres and in panel (b) for cluster edges.

The dashed curve in Fig.~\ref{fig:lgt-lgm}(a,b) presents a 
``mass--MS-lifetime relation'' i.e., a theoretical scale taken at the TAMS
from the Padova isochrones with overshooting for $Z=0.019$. The relation sets an upper
limit of masses of MS stars which can be expected in a cluster of a given age.
In cluster centres, the most massive members evolve in good agreement with the
``mass--MS-lifetime relation'' (Fig.~\ref{fig:lgt-lgm}a), whereas this evolution
is rather weak at the cluster edges (Fig.~\ref{fig:lgt-lgm}b).
In absence of mass segregation equations (\ref{eqn:lgm-t0}) and
(\ref{eqn:lgm-t1}), which are presented by solid lines in Fig.~\ref{fig:lgt-lgm}(a,b),
should show up similar coefficients. This, however, is not the case.
A large difference between masses in the central and outer regions of clusters
at earlier ages indicates a strong mass segregation, on average. 
Due to the burning out of massive stars in the cluster cores (Fig.~\ref{fig:lgt-lgm}a),
the mass difference is decreasing with cluster age and approaches zero
at $\log t > 9$. Nevertheless, one can assume that there should be members
with masses lower than about $1\,m_\odot$ in the outer regions of very old clusters, but
the relatively bright limiting magnitude of the \ascc prevents us from observing
them already at $(V-M_{V}) > $ 7. 
 
Variations of mass segregation with age observed in our cluster sample can 
be quantitatively analysed from the radial mass gradient computed by 
equation (\ref{eqn:lgm-rr})
and presented in Fig.~\ref{fig:lgt-lgm}c. A strongly negative mass gradient points out
a general concentration of the most massive members to the cluster centres,
whereas a large individual $rms$ error of a mass gradient indicates mainly that  
relatively massive stars are present also outside the very central region 
(different situations are illustrated in Fig.~\ref{fig:rr-lgm-p}).

Although, a number of clusters do not show indication for mass segregation,
a systematic trend of $d\log m_{max}/dr$ towards negative values can be seen 
over the whole range of cluster ages in Fig.~\ref{fig:lgt-lgm}c.
At $\log t>$ 7.6 the radial mass gradient flattens steadily, from about $-0.4$ to
$-0.1$ at $\log t \approx$ 8.9. As we discussed above (Fig.~\ref{fig:lgt-lgm}a), 
this flattening can be explained by the gradual evolution of the
most massive stars  away from the MS in the central areas of clusters.
Nearby clusters clearly support this general trend though their average curve 
is shifted to smaller mass gradients: in these clusters we are 
able to observe stars with masses slightly below $1\, m_{\odot}$ which are 
widely distributed over the cluster area, up to the cluster edge 
(cf. Fig.~\ref{fig:lgt-lgm}b). Therefore, a stronger mass
segregation can be still seen in nearby clusters with 8.5 $<\log t<$ 9.0. 

The scattering of data points in Fig.~\ref{fig:lgt-lgm}c indicates a dependence on 
cluster ages. The standard deviation of the mass gradient is 0.10-0.15 for
clusters with $\log t>$ 7.5 and it is larger by a factor of two (0.25-0.30)
for younger clusters. According to an F-test, the hypothesis of equal variances
is clearly rejected for clusters younger and older than $\log t=$ 7.5. Also, individual
$rms$ errors of $d\log m_{max}/dr$ are, on average, twice as large for
younger clusters as for older ones. Taking this into account, we conclude that
the group of clusters with $\log t<$ 7.5 is less homogeneous than the group of 
older clusters, and that the apparent distribution of the most massive stars over the area
cannot always be described sufficiently well in young clusters by 
equation (\ref{eqn:lgm-rr}). 

A possible explanation is that our sample at $\log t\lesssim7.5$ presents
a mixture of young clusters with a different grade of mass segregation.
Some of them have a significant negative mass gradient (e.g., ASCC 58 in 
Fig.~\ref{fig:rr-lgm-p}) supporting the conclusion that substantial mass segregation has 
occurred here already at early stages of the evolution. The absence of clusters
younger than 5 Myr in our sample prevents us, however, to understand whether
the mass segregation has a primordial character (i.e. originated during cluster
formation) or if it is already a result of the dynamical evolution during the first
5 Myr of the cluster's life. In contrast, a few other clusters 
(e.g., Cr 121 in Fig.~\ref{fig:rr-lgm-p}) have a flatter radial mass gradient, 
not significantly differing from zero. In these clusters, the most
massive stars show a stochastic distribution over the cluster area, and 
it looks as if they are hampered in their dynamical evolution.
Later, at about $\log t=7.7...8.0$ the majority of clusters seems
to achieve a quasi-equilibrium,
and the evolution of massive stars from the MS becomes a more prominent
effect in the observations.

Among 31 clusters with $\log t<$ 7.5, our sample contains 10 clusters
with significant negative mass gradients ($d\log m_{max}/dr + 3\sigma < 0$
where $\sigma$ is the individual $rms$ error of the radial mass gradient)
and 6 clusters with mass gradients not significantly differing from zero
(|$d\log m_{max}/dr$| $<\sigma$). Nevertheless, the sampling is rather poor
and the scattering of data points is too large to 
make a  statistically significant conclusion. Also, we cannot exclude the possibility that
a low significance of the mass gradient can be a consequence of uncertainties in
the determination of the cluster centres due to an irregular and patchy distribution of
the absorbing matter within young clusters.

We conclude that the observed relation between the radial mass gradient and the age
of a cluster can be interpreted in terms of the dynamical and stellar
evolutions. Our sample does not include very young clusters, but
at least at an age of 5...10 Myr a strong radial concentration of massive
stars can be already observed in several clusters. Numerical simulations
(Khalisi et al.~\cite{kspur}) indicate that, depending on the
initial mass of a cluster, mass segregation can occur very rapidly
for massive members. On the other hand,
this process can be hampered by stellar winds, ionisation fronts etc.
of the most massive stars, especially if a cluster is located
within a large star forming region (Kroupa et al.~\cite{krea01}). This is
possibly the reason why we observe clusters of the same ages, where
some show mass segregation and other do not show it at all.
Nevertheless, during the following 50...100~Myr the
dynamical evolution takes overhand due to pair encounters and energy equipartition,
and we observe a more regular pattern in the distribution of stars
of different masses. This age seems to be typical for the relaxation
of clusters in our sample. 
%At least, no significant development of mass segregation for
%stars of moderate masses ($m\sim 1-2\,m_\odot$) can be concluded from the distribution
%of $d\log m_{max}/dr$ at $\log t>$ 8 though, we should not exclude that the energy 
%equipartition due to pair 
%encounters may be still effective for stars with masses $m<1\,m_{\odot}$ i.e.,
%below the limit of our observations. 
Although mass segregation can still continue at lower masses, the apparent mass distribution in
clusters of our sample at $\log t>$ 8 is mainly
governed by stellar evolution removing the most massive stars from
the {}``scene''. External gravitational shocks may also influence the mass
distribution in clusters and can be partly responsible for a spread of
the radial mass gradient at $\log t > 8$.

\section{Summary\label{sec:concl}}

This study is based on the Catalogue of Open Cluster Data (\clucat)
and its Extension~1 described in Papers II and III. The \clucat is derived
from the \ascc,
a homogeneous all-sky catalogue with complete information on proper motions and
$B,V$-photometry.
So, all open clusters found in this
catalogue can be treated in the same way to derive their astrophysical parameters.
On the other hand, the price to be paid for this advantage
is the bright completeness limit of \ascc at about $V$~=~11.5.
However, the
biases resulting from a simply magnitude limited sample can be estimated,
they have been discussed in the previous sections and
have been taken care of in order not to influence the conclusions.
Using samples of clusters from different sources with different photometry
and/or different limiting magnitude may introduce biases in the results
which cannot be estimated easily.

The whole sample from \ascc consists of 641 open clusters.
In papers II and III we determined membership in the clusters applying
photometric as well as astrometric criteria.
Apparent linear radii have been computed from individual distances and angular
sizes of the clusters, based on members only. For the first time, 
the structural properties of the galactic open cluster system have been 
statistically analysed from an unbiased, homogeneous, and  
relatively large sample.
A comparison of our cluster sizes
with those given in Lyng\aa{}~(\cite{lyn87}) (about 500
clusters in common) shows that cluster radii from Lyng\aa{}~
are in average lower by a factor of 2, and they fit rather the core
than the corona.

Our large sample allowed us to investigate the dependence of the cluster
size on the age of a cluster and on its location in the Galaxy.
The clusters cover an age range between about 5~Myr to more than 1~Gyr.
For younger clusters ($<$ 200 Myr) there is no significant
correlation between linear size and Galactocentric distance.
At an age corresponding to two revolutions around
the Galactic centre we detect that the clusters are
on average smaller ($\overline{R_{cl}}=3.8\pm0.2$~pc) inside the solar
circle than outside ($\overline{R_{cl}}=4.6\pm0.3$~pc).
According to a ($K-S$) test the probability that both subsamples are
drawn from the same distribution is less than 4~\%.
This size
dependence on Galactocentric radius lead to the conclusion that the inner
Galactic disk is void with respect to older open clusters.
No clusters older than the age of the Hyades should exist inside a Galactocentric radius
of about 6~kpc.
Perpendicular to the plane we note
a systematic increase of cluster
sizes with increasing $|Z^{\prime}|$. This, however,
turned out to be significant only for
clusters older than $\log t>8.35$, which already survived at least
one revolution around the Galactic centre.

From these findings the following picture of
the evolution of open clusters arises.
Clusters in the wider Solar neighbourhood
are formed within the thin disk,
their initial size distribution does not show a significant
correlation with the $R_{G}$-- and $|Z^{\prime}|$- coordinates.
The size distribution changes at ages corresponding to one revolution around the Galactic centre.
At low $Z^{\prime}$ we now note a relatively larger number of small clusters.
This makes us conclude that close to the Galactic equator and
inside the solar circle larger clusters are in danger to dissolve
even during the first revolution around the Galactic centre.
On the other hand, they have a higher chance to survive encounters
and the impact of Galactic tidal forces,
if their orbits are outside the Solar one and are inclined
to the Galactic plane. Therefore, they reach higher ages at these locations.
Finally, the apparent linear
sizes of clusters and their cores are, on average, decreasing with time and
this process is faster for the coronae than for
the cores. Taking into account that our input catalogue is magnitude limited,
this finding can be interpreted as a first hint for mass segregation.

In the majority of clusters of our sample clear evidence for mass segregation 
of stars with $m>1\,m_\odot$ has been 
established from the distribution of the radial mass gradient as a function 
of age. 
An apparent flattening of the radial mass gradient for clusters older than 50...100 Myr
occurs due to stellar
evolution when massive stars subsequently leave the main sequence, and, secondly,
because we cannot observe the low-mass stars due to the bright limiting
magnitude of the \ascc.
External gravitational shocks may also influence the mass
distribution in clusters and can be partly responsible for a spread of
the radial mass gradient at $\log t > 8$. Nevertheless, a ``typical''
cluster older than about 100 Myr and within about 1~kpc from the Sun shows 
mass segregation.

The youngest clusters of our sample with ages less than 50 Myr show a large
spread of the radial mass gradient: from clusters
with a clear concentration of the most massive stars to the centres 
up to clusters with no or only a flat mass gradient. The different dynamical state
of clusters of the same age possibly results from the different initial conditions and 
environments of the clusters. 

\begin{acknowledgements}
This work was supported by the DFG grant 436~RUS~113/757/0-1, and
RFBR grant 06-02-16379. We acknowledge a discussion of the
results with Rainer Spurzem. We are grateful to the anonymous referee
for his/her useful comments.
\end{acknowledgements}

\end{document}